\documentclass[11pt]{article}
\usepackage{amssymb}
\usepackage{amsmath,amsthm}  

\usepackage{ulem} 
\renewcommand{\em}{\it}

\usepackage{graphicx}
\usepackage{lmodern}
\usepackage{microtype}
\usepackage{pstricks}

\usepackage[english,francais]{babel}

\newfont{\ensmathquatorze}{msbm10 scaled 1400}
\newfont{\ensmathonze}{msbm10 scaled 1100}
\newfont{\ensmathdix}{msbm10}
\newfont{\ensmathneuf}{msbm10 scaled 833}
\newfont{\ensmathhuit}{msbm10 scaled 694}
\newfam\ensmathfam                        
\textfont\ensmathfam=\ensmathonze        
\scriptfont\ensmathfam=\ensmathdix       
\scriptscriptfont\ensmathfam=\ensmathhuit
\def\ensmf{\fam\ensmathfam\ensmathonze}         

\renewcommand{\leq}{\leqslant}
\renewcommand{\geq}{\geqslant}

\def\eqdef{\stackrel{\mbox{\tiny def}}{=}}     
\def\eqlaw{\stackrel{\mbox{\tiny (law)}}{=}}     
\newcommand{\ket}[1]{|\kern.3ex#1\kern.3ex\rangle}
\newcommand{\bra}[1]{\langle\kern.3ex #1 \kern.3ex|}

\newcommand{\mean}[1]{\left\langle #1 \right\rangle} 
\newcommand{\smean}[1]{\langle #1 \rangle} 

\newcommand{\EXP}[1]{{\mbox{\large e}}^{#1}}         

\newcommand{\argsinh}{\mathop{\mathrm{argsinh}}\nolimits}

\newcommand{\im}{\mathop{\mathrm{Im}}\nolimits}      
\newcommand{\cotg}{\mathop{\mathrm{cotg}}\nolimits}  
  
\renewcommand{\max}[2]{\mathop{\mathrm{max}}\nolimits\left( #1 , #2\right)}

\def\NN{{\ensmf N}}                 
\def\RR{{\ensmf R}}                 

\def\I{{\rm i}}                  
\def\D{{\rm d}}                  

\newcommand\antiddots{\mathinner{\mkern2mu\raise1pt\hbox{.}\mkern2mu
\newline \raise4pt\hbox{.}\mkern2mu\raise7pt\hbox{.}\mkern1mu}}

\begin{document}

\selectlanguage{english}

\title{Breaking supersymmetry in a \\ one-dimensional random Hamiltonian}

\author{ Christian Hagendorf$^{(a)}$ and Christophe Texier$^{(b,c)}$ }

\date{August 29, 2008}

\maketitle

\hspace{1cm}
$^{(a)}$ 
\begin{minipage}[t]{13cm}
{\small
Laboratoire de Physique Th\'eorique de l'\'Ecole Normale Sup\'erieure,

24, rue Lhomond, F-75230 Paris Cedex 05, France.
}
\end{minipage}

\vspace{0.25cm}

\hspace{1cm}
$^{(b)}$
\begin{minipage}[t]{13cm}
{\small
 Laboratoire de Physique Th\'eorique et Mod\`eles Statistiques,
UMR 8626 du CNRS, 

Universit\'e Paris-Sud, B\^at. 100, F-91405 Orsay Cedex, France.
}
\end{minipage}

\vspace{0.25cm}

\hspace{1cm}
$^{(c)}$\begin{minipage}[t]{13cm}
{\small
 Laboratoire de Physique des Solides, UMR 8502 du CNRS, 

Universit\'e Paris-Sud, B\^at. 510, F-91405 Orsay Cedex, France.
}
\end{minipage}

\begin{abstract}
  The one-dimensional supersymmetric random Hamiltonian
  $H_\mathrm{susy}=-\frac{\D^2}{\D{}x^2}+\phi^2+\phi'$, where $\phi(x)$ is a
  Gaussian white noise of zero mean and variance $g$, presents particular
  spectral and localization properties at low energy~: a Dyson singularity in
  the integrated density of states (IDoS) $N(E)\sim1/\ln^2E$ and a
  delocalization transition related to the behaviour of the Lyapunov exponent
  (inverse localization length) vanishing like $\gamma(E)\sim1/|\ln{}E|$ as
  $E\to0$.
  We study how this picture is affected by breaking supersymmetry with
  a scalar random potential~: $H=H_\mathrm{susy}+V(x)$ where $V(x)$ is
  a Gaussian white noise of variance $\sigma$. In the limit
  $\sigma\ll{g}^3$, a fraction of states
  $N(0)\sim{g}/\ln^2(g^3/\sigma)$ migrate to the negative spectrum and
  the Lyapunov exponent reaches a finite value
  $\gamma(0)\sim{g}/\ln(g^3/\sigma)$ at $E=0$.
  Exponential (Lifshits) tail of the IDoS for $E\to-\infty$ is studied
  in detail and is shown to involve a competition between the two
  noises $\phi(x)$ and $V(x)$ whatever the larger is.
  This analysis relies on analytic results for $N(E)$ and $\gamma(E)$
  obtained by two different methods~: a stochastic method and the replica
  method.
  The problem of extreme value statistics of eigenvalues is also
  considered (distribution of the $n-$th excited state energy).
  The results are analyzed in the context of classical diffusion in a random
  force field in the presence of random annihilation/creation local rates.
\end{abstract}

\section{Introduction}

The study of spectral and localization properties of one-dimensional (1d)
random Hamiltonians has stimulated a huge activity since the pioneering
works of Dyson~\cite{Dys53}, Schmidt~\cite{Sch57}, Frisch \&
Lloyd~\cite{FriLlo60}, Lifshits~\cite{Lif65} and many
others (references to the most important works may be found in the
review article \cite{Gog82} and the books \cite{LifGrePas88,Luc92}).
Because the dimension plays an important role in localization
problems~\cite{AbrAndLicRam79}, the strictly one-dimensional situation misses
some features of higher dimension case (like a weak localization regime). On
the other hand the one-dimensional case allows to make use of powerful
nonperturbative methods and study subtle properties which are much more difficult to
tackle in higher dimensions.
The random Schr\"odinger Hamiltonians
$H_\mathrm{scalar}=-\frac{\D^2}{\D{}x^2}+V(x)$, where $V(x)$ is a
random function, have been studied in great detail~\cite{LifGrePas88}
and their properties are rather generic under the asumption that
$V(x)$ is correlated on a small length scale and
$\int\D{x}\,\smean{V(x)V(0)}$ remains finite\footnote{
  Some interesting results have been also obtained in
  Ref.~\cite{Luc05} in a situation where the correlation function growths
  at large distance like $\smean{V(x)V(0)}\sim|x|^\eta$ with $\eta>0$
  (the case $\eta=1$ corresponds to  a Brownian motion).
}~: exponential tail in the density of states\footnote{
  The form of the exponential Lifshits tail depends on the details of the
  distribution of the random potential. Note that the spectrum of the
  Hamiltonian $H_\mathrm{scalar}=-\frac{\D^2}{\D{}x^2}+V(x)$ can also presents power-law
  singularity~: for a random potential describing a weak concentration of
  impurities of negative weights, each trapping a localized state at energy
  $E_0<0$, the spectrum presents a power law singularity near $E_0$, with an
  exponent proportional to the concentration of impurities~; such a
  singularity is called a Halperin
  singularity~\cite{Sch57,FriLlo60,Lif63,BycDyk66a,Hal67}).
}
at low energies (Lifshits
singularity)~\cite{Sch57,LaxPhi58,Hal65,FriLlo60,BycDyk66a,AntPasSly81,LifGrePas88}
and decreasing Lyapunov exponent (inverse localization length) at high
energy~\cite{AntPasSly81,LifGrePas88} $\gamma\propto1/E$ for $E\to+\infty$.
The situation can be quite different if the Hamiltonian possesses some
symmetry preserved by the introduction of the random potential. Such
a situation occurs in the case of supersymmetric random Hamiltonian
\begin{equation}
  \label{Hsusy}
  H_\mathrm{susy} = -\frac{\D^2}{\D x^2} + \phi(x)^2 + \phi'(x) 
\end{equation}
This Hamiltonian has a positive spectrum, a direct consequence of the fact
that it can be factorized in the form $H_\mathrm{susy}=Q^\dagger{}Q$ with
$Q=-\frac{\D}{\D x}+\phi(x)$ and $Q^\dagger=\frac{\D}{\D x}+\phi(x)$.
Moreover, it is worth pointing out that
$H_\mathrm{susy}\equiv{H_+}=Q^\dagger{}Q$ and its supersymmetric partner
$H_-=QQ^\dagger=-\frac{\D^2}{\D{}x^2}+\phi^2-\phi'$ are the two components of the square of
the Dirac Hamiltonian $\mathcal{H}_D=\sigma_2\I\frac{\D}{\D{}x}+\sigma_1\phi(x)$, where
$\sigma_i$ are Pauli matrices~:
$\mathcal{H}_D^2=-\frac{\D^2}{\D{}x^2}+\phi^2+\sigma_3\phi'$. Therefore the Hamiltonian
(\ref{Hsusy}) arises naturally when studying random Dirac Hamiltonians.
Besides its own interest for the physics of localization, this model
is relevant in several physical contexts like classical diffusion in a
random force field (Sinai
problem)~\cite{Sin82,BouComGeoLeD87,BouComGeoLeD90,Mon95,FisLeDMon98,LeDMonFis99}
(see\ \S~\ref{sec:motiv} below), organic
conductors~\cite{TakLinMak80,Mon95} or spin chains (the spectrum of
excitations of an antiferromagnetic spin-chain is linear at small
energies like in free fermion model~; the precise mapping of AF
spin-chain to free fermions can be achieved thanks to a Jordan-Wigner
transformation)~\cite{Zim82,McK96,FabMel97,BalFis97,SteFabGog98,FisLeDMon98}~;
see the review provided in Ref.~\cite{BouComGeoLeD90}.
The relation to discrete models has been discussed~: the
supersymmetric Hamiltonian is the continuum limit of a discrete
tight-binding Hamiltonian with off-diagonal disorder~\cite{Mon95}. It
is also the continuum limit of a tight-binding Hamiltonian with
diagonal disorder at the band center~\cite{OvcEri77,LifGrePas88} (this
point has been recently rediscussed in \cite{SchTit03}). The
supersymmetry is responsible for rather particular spectral and
localization properties. For the sake of concreteness, let us choose
for $\phi(x)$ a Gaussian white noise of zero mean, $\mean{\phi(x)}=0$
and $\mean{\phi(x)\phi(x')}=g\,\delta(x-x')$. In the low energy limit,
$E\ll{}g^2$, the integrated density of states (IDoS) presents the
Dyson singularity
$N(E)\simeq2g/\ln^2(g^2/E)$~\cite{GogMel77,OvcEri77,BouComGeoLeD87,BouComGeoLeD90},
similar to the one of the spring chain with random masses~\cite{Dys53}
or Anderson model with random hoppings~\cite{TheCoh76,EggRie78,Zim82}.
The Lyapunov exponent vanishes as
$\gamma(E)\simeq2g/\ln(g^2/E)$~\cite{BouComGeoLeD90} (also obtained for
discrete models in Refs.~\cite{TheCoh76,EggRie78,Zim82}), indicating a
delocalization transition.  This delocalization transition is suported
by studying other quantities~: ({\it i}) Statistical properties of the
zero mode wave function \cite{BroKre95,SheTsv98,ComTex98} indicate
long range power law correlations (like the Lyapunov exponent
analysis, these calculations do not account for boundary conditions).
({\it ii}) The distribution of the transmission probability through a
finite slab of length $L$ at zero energy.  In particular the average
transmission decreases like $1/\sqrt{L}$~\cite{SteCheFabGog99}, that
is slower than the behaviour $1/L$ for a quasi 1d conducting weakly
disordered wire.  ({\it iii}) Time delay distribution presents a
log-normal distribution at zero energy~\cite{SteCheFabGog99,Tex99}.
({\it iv}) The conductivity is found to be finite at $E=0$
\cite{GogMel77,Gog82}.  ({\it v}) Finally, the study of extreme value
statistics of energy levels indicates spectral correlations for
$E\to0$~\cite{Tex00}.  In the high energy limit $E\to\infty$, the
localization properties are quite unusual since the Lyapunov exponent
does not vanish but reaches a finite value
$\gamma(E\to\infty)\simeq{g}/2$. This property is due to the singular
nature of the potential $\phi^2+\phi'$ with $\phi$ a white noise. When
the potential is regularized by introducing a small but finite
correlation length, it has been shown in \cite{ComDesMon95,Mon95} that
the Lyapunov exponent decreases as $\gamma\propto1/E$ for largest
energies, as for the random Hamiltonian
$H_\mathrm{scalar}=-\frac{\D^2}{\D{}x^2}+V(x)$.
If the random function $\phi(x)$ possesses a finite mean value
$\mean{\phi(x)}=\mu{g}$, logarithmic singularities are converted into power
law singularities~\cite{OvcEri77,BouComGeoLeD90}.
Extension to more general situations has been considered in
Ref.~\cite{Boc99}, where spectrum and localization have been studied
for the most general random Dirac 1d Hamiltonian (random mass, random
scalar field, random gauge field), however such a study still
preserves the (particle-hole) symmetry of the Hamiltonian (note that
the distribution of the local DoS for this model has been investigated
in Ref.~\cite{BunMcK01}).

The aim of the present article is to discuss the effect of the addition of a
scalar random potential that breaks the supersymmetry~:
\begin{equation}
  \label{eqn:hamiltonian}
  \boxed{  H = -\frac{\D^2}{\D x^2} + \phi(x)^2 + \phi'(x) + V(x) }
\end{equation}
We will mostly consider the case when the functions $\phi$ and $V$ are two
uncorrelated Gaussian white noises with variances
$\mean{\phi(x)\phi(x')}=g\,\delta(x-x')$ and
$\mean{V(x)V(x')}=\sigma\,\delta(x-x')$. The case with a finite
$\mean{\phi(x)}$ will be studied in the section~\ref{sec:replica}
with  the replica
method. The case of correlated Gaussian white noises $\phi$ and $V$
will be discussed in the appendix~\ref{app:corrnoises} where it is
mapped onto the problem of uncorrelated noises. 
Our purpose is to study how spectral and localization properties of
$H_\mathrm{susy}$ are modified when introducing the scalar potential.
A first obvious change is that the spectrum of $H$ is not restricted
to be positive. Natural questions are therefore~: what is the number
of states sent to $\RR^-$ by the introduction of the potential $V(x)$,
how their energies are distributed~? How the delocalization at
$E\to0$ for the Hamiltonian $H_\mathrm{susy}$ is affected~?

The paper is organized as follows. After giving a physical motivation for our
model right hereafter, we study spectral and localization properties of $H$ in
sections \ref{sec:spec} and \ref{sec:loc} respectively. Our approach relies on
well-established techniques of stochastic differential equations. In section
\ref{sec:replica}, we employ the replica method in order to find other
analytical expressions for the IDoS and the Lyapunov exponent and consider the
more general case of a finite~$\mean{\phi(x)}$.

\subsection{A motivation~: branching random walks in a disordered environment\label{sec:motiv}}

Let us first recall the well-known relation between the Fokker-Planck equation
(FPE) describing classical diffusion in a force field $\phi(x)$ and the
Schr\"odinger equation for a potential $\phi^2+\phi'$. Let us consider the
Langevin equation $\frac{\D x(t)}{\D t}=2\phi(x(t))+\sqrt2\,\eta(t)$, where
the Langevin force $\eta(t)$ is a normalized white noise. This equation is
related to the FPE $\partial_tP(x;t)=F_xP(x;t)$ where the forward generator
reads $F_x=\partial_x^2-2\partial_x\phi(x)$. The FPE can be transformed into
the Schr\"odinger equation $-\partial_t\psi(x;t)=H_\mathrm{susy}\psi(x;t)$
thanks to the nonunitary transformation $P(x;t)=\psi_0(x)\psi(x;t)$ since
\begin{equation}
  \label{isospectransf}
  \psi_0(x)^{-1}F_x\psi_0(x)=-H_\mathrm{susy}
  \hspace{0.5cm}\mbox{where}\hspace{0.5cm}
  \psi_0(x)=\EXP{\int^x\D{x}'\phi(x')}
\end{equation}
Note that the operator transformation $F_x\to{}H_\mathrm{susy}$ is
isospectral. $\psi_0(x)$ is annihilated by the operator $Q$ defined
above~:
$Q\psi_0=0$.  For a
confining force field, $\psi_0(x)$ is the normalizable zero mode of
$H_\mathrm{susy}$ and is related to the stationary distribution of the
FPE~: $P(x;t\to\infty)\simeq\psi_0(x)^2$.

In order to propose the physical interpretation of the last term of
(\ref{eqn:hamiltonian}), we start from a discrete formulation of the problem
of diffusion-controlled reaction in a one-dimensional quenched random
potential landscape $\mathcal{V}_k$. Let us consider non-interacting particles
on an infinite one-dimensional lattice with lattice spacing $a$. We label
lattices site by $k\in\mathbb{Z}$, corresponding to a position $ka$. We
allow the local occupation number $n_k$ for site $k$ to take arbitrary
positive integer values (bosonic particles). The transition rates between
neighbouring sites $k$ and $k+1$ can be obtained from the Arrhenius law
\begin{equation}
  t_{k+1,k} = \frac{1}{a^2}\,\EXP{\mathcal{V}_k-\mathcal{V}_{k+1}}
  \label{eqn:rates}
\end{equation}
where $\mathcal{V}_k$ is the potential at site $k$. The prefactor is chosen in
order to obtain a well-defined continuum limit $a\to 0^+$. Additionnally we
consider the following chemical reactions~: we allow particle replication
$\text{A}\to m\text{A}$, $m\geq 2$, with a local rate $\beta_{m,k}$ and
particle annihilation $\text{A}\to\emptyset$ with a local rate $\gamma_k$. The
reaction rates are supposed to be random quantities. Therefore, the model
describes branching random walks in a one-dimensional disordered environment,
including particle annihilation.

Let us study the particle distribution on the lattice~: we denote $n_k$ the
occupation of site $k$. Its mean value obeys the following master equation
\begin{equation}
  \frac{\D\overline{n}_k}{\D t}= 
  t_{k,k+1}\, \overline{n}_{k+1} + t_{k,k-1}\, \overline{n}_{k-1} - 
  (t_{k+1,k}+t_{k-1,k})\,\overline{n}_k + (\beta_k-\gamma_k)\, \overline{n}_k
  \label{eqn:density}
\end{equation}
where averaging $\overline{\cdots}$ is taken with respect to the random
dynamics defined by rates (\ref{eqn:rates}) (not to be confused with 
averaging $\smean{\cdots}$ with respect to the quenched random
potential $\mathcal{V}_k$ and
random annihilation/creation rates). We have introduced
$\beta_k=\sum_{m=1}^\infty{}m\beta_{m+1,k}$.
For the continuum limit, we introduce the density
$n(x=ka,t)=\overline{n}_k/a$. As $a\to 0$ we develop
$\frac1a\overline{n}_{k\pm1}=n(x,t)\pm{}a\,\partial_xn(x,t)+\frac12a^2\,\partial_x^2n(x,t)+\cdots$
Moreover, we introduce the force field $\phi(x)$ via
$\mathcal{V}_k-\mathcal{V}_{k+1}=a\phi(x=ka)+\frac12a^2\phi'(x=ka)+\cdots$
what allows us to develop the transitions rates (\ref{eqn:rates}) as
\begin{align}
  t_{k,k\pm1}=\frac{1}{a^2} \mp \frac{\phi(x)}{a} 
  -\frac{\phi'(x)}{2}+\frac{\phi(x)^2}{2}+\cdots,\qquad
  t_{k\pm1,k} = 
  \frac{1}{a^2}\pm \frac{\phi(x)}{a}+\frac{\phi'(x)}{2}+\frac{\phi(x)^2}{2}+\cdots
\end{align}
We also introduce the notation $\gamma_k-\beta_k=V(x=ka)$ for the difference
of annihilation rates and creation rates ($V(x)>0$ corresponds to annihilation
and $V(x)<0$ to creation). The development yields the partial differential
equation
\begin{equation}
  \frac{\partial n(x,t)}{\partial t} = 
  \frac{\partial^2 n(x,t)}{\partial x^2}  -
  2\frac{\partial}{\partial x}[\phi(x)\,n(x,t)] - V(x)\,n(x,t) 
  = -H_\mathrm{FP}n(x,t)
\end{equation}
for the average particle density, with $H_\mathrm{FP}=-F_x+V(x)$. We will
consider the case where the random force field $\phi(x)$ and the random
annihilation/creation rates $V(x)$ are correlated over small scale. For large scale
properties of the diffusion, the minimal model corresponds to assume that
$\phi(x)$ and $V(x)$ are two Gaussian white noises. The mean value
$\smean{\phi(x)}$ corresponds to the average drift of particles and
$\smean{V(x)}$ is related to the average rate of particle annihilation at $x$.
We will first consider the case $\smean{\phi(x)}=0$ (the case of finite drift
will be discussed in section~\ref{sec:replica}). A finite average creation
rate $\smean{V(x)}$ corresponds to a trivial global shift of the spectrum of
$H$, therefore we will set~$\smean{V(x)}=0$.

We have introduced a Fokker-Planck-like differential operator
$H_{\mathrm{FP}}$ which, as explained above, may be related to the
Schr\"odinger operator (\ref{eqn:hamiltonian}) thanks to the isospectral
transformation (\ref{isospectransf})~: $\psi(x,t)=\psi_0(x)n(x,t)$.
Hence, the spectrum of $H$ is of great interest for the diffusion problem. In
particular, if we wish to determine the density $n(x,t|y,0)$ with initial
condition $n(x,0|y,0)=\delta(x-y)$ we may rewrite in terms of the spectrum
$\{E_\alpha,\,\Psi_\alpha(x)\}$ of~$H$
\begin{equation}
  n(x,t|y,0) =
  \frac{\psi_0(x)}{\psi_0(y)}
  \sum_\alpha \Psi_\alpha(x)\,\Psi_\alpha(y) \, \EXP{-E_\alpha t}
\end{equation}
where $\psi_0(x)$ is the zero mode of $H_\mathrm{susy}$ given above.
A first quantity to consider is the average occupation at $x$ at time
$t$ after release of a particle at $y=x$ at time $t=0$. We can use the
translation invariance of the problem  to identify the position average
with averaging with respect to disorder~:
\begin{equation}
  \label{Laplace}
   \smean{n(x,t|x,0)}
  =\lim_{L\to\infty}\frac{1}{L}\int_{-L/2}^{+L/2}\D{x}\,n(x,t|x,0) 
  =\int_{-\infty}^{+\infty} \D E\,\rho(E)\,\EXP{-Et}
\end{equation}
where $\rho(E)$ denotes the density of states of $H$ (we have omitted
averaging in the {\it r.h.s} thanks to self averaging
properties of the density of states).
This relation shows that low energy properties of the quantum Hamiltonian are
related to large time asymptotics for the return probability of the classical
diffusion problem.

\section{Spectral properties}
\label{sec:spec}

In this section we recall the phase formalism, the continuous version
of the well-known Dyson-Schmidt method~\cite{Dys53,Sch57,Luc92}. A
clear presentation can be found in
Refs.~\cite{AntPasSly81,LifGrePas88}. The basic idea relates on the
Sturm-Liouville theorem stating that the number of nodes of the
one-dimensional wavefunction of energy $E$ is equal to the number of normalizable
states below $E$. The starting point is to convert the Sturm-Liouville
problem\footnote{
  A spectral problem is formulated as~: 
  find the solutions of $H\psi(x)=E\psi(x)$ for
  some boundary conditions, e.g. $\psi(0)=\psi(L)=0$. On a finite
  interval, such solutions $(\psi_n(x),\,E_n)$, exist only for discrete
  values of the energy $E\in\mathrm{Spec}(H)=\{E_n\}$.
} 
into a Cauchy problem\footnote{
  Solve $H\psi(x;E)=E\psi(x;E)$ for given intial conditions, e.g.
  $\psi(0;E)=0$ and $\psi'(0;E)=1$. Solutions exist~$\forall\,E$.
} 
and study the statistical properties of the solution of this latter problem.
The next step consists to separate the solution into an oscillating part and
an envelope $\psi(x;E)=\rho_E(x)\,\sin\theta_E(x)$. The study of the phase
$\theta_E(x)$ permits to analyze the spectral properties of the Hamiltonian
$H$ since it allows to count the number of nodes of the wave function. The
damping of the envelope characterizes its localization properties. Strictly
speaking, $\psi(x;E)$ is the wavefunction only if $E$ coincides with an
eigenvalue $\psi(x;E_n)\propto\varphi_n(x)$, that is when the second boundary
condition is satisfied $\psi(x=L;E_n)=0$.

\subsection{Ricatti variable}
\label{sec:ricatti}

It is convenient to start by introducing the ``Ricatti'' variable
$z\eqdef\psi'/\psi-\phi$, the Schr\"odinger equation $H\psi=E\psi$ leads to the
stochastic differential equation (SDE)~:
\begin{equation}
  \label{1stSDE}
  \frac{\D}{\D x} z(x) = - E - z(x)^2 - 2\,z(x)\,\phi(x) + V(x)
  \hspace{1cm}\mbox{  (Stratonovich)}
\end{equation}
Since the random functions $\phi$ and $V$ are understood to be the white noise
limits of some physical regular noises (correlated over a finite length scale),
the SDE must be understood in the Stratonovich sense~\cite{Gar89}. The
relation (\ref{theorem}) derived in appendix \ref{sec:usefulth} allows
to simplify (\ref{1stSDE}) in order to deal with one noise only
\begin{equation}
  \label{SDEZ}
  \D z \eqlaw  -(E +z^2)\,\D x + \sqrt{\sigma+4gz^2}\,\D W(x)
  \hspace{1cm}\mbox{  (Stratonovich)}
\end{equation}
where $W(x)$ is a normalized Wiener process (primitive of a white noise). We
define $\beta(z)=\sqrt{\sigma+4gz^2}$. This Langevin equation is related to a
Fokker-Planck equation (FPE) $\partial_xT(z;x)=F_zT(z;x)$ where
$F_z=\partial_z(E+z^2)+\frac12[\partial_z\beta(z)]^2$ is the forward
generator. This equation admits a stationary solution for a constant flow. The
current of $z$ through $\RR$ corresponds to the number of divergencies of the
Ricatti variable per unit length, therefore to the number of zeros of the wave
function per unit length. This is precisely the average integrated density of
states (IDoS) per unit length $N(E)$. Therefore
\begin{equation}
  \label{eqT}
  N(E) = (z^2+E)\,T(z) + \frac12\beta(z)\,\frac{\D}{\D z}[\beta(z)T(z)]
\end{equation}
We recover on this particular case the general Rice formula
$\lim_{z\to\infty}z^2T(z)=N(E)$. We introduce the function
$\mathcal{U}(z)=4g\int^z_0\D{z'}\,\frac{E+z'^2}{\beta(z')^2}$,
\begin{equation}
  \mathcal{U}(z) = z + \sqrt{\frac{\sigma}{4g}}
  \left(\frac{4Eg}{\sigma}-1\right)
  \arctan\left(\sqrt{\frac{4g}{\sigma}}\,z\right)
\end{equation}
we obtain the distribution~:
\begin{equation}
  \label{disRicatti}
  T(z) = \frac{2N(E)}{\beta(z)} \EXP{-\frac1{2g}\mathcal{U}(z)}
  \int_{-\infty}^{z}\frac{\D{z'}}{\beta(z')}\,\EXP{\frac1{2g}\mathcal{U}(z')}
\end{equation}
Imposing normalization gives an explicit expression of the IDoS.

\subsection{Phase and envelope}

The phase formalism introduces another set of variables that give a more
transparent picture to analyze spectrum and localization.

\hspace{0.25cm}

\noindent{\bf Positive part of the spectrum~:}
\mathversion{bold}$E=+k^2$.--\mathversion{normal} We write
$H_\mathrm{susy}=Q^\dagger{}Q$ with $Q=-\frac{\D}{\D x}+\phi(x)$ and
$Q^\dagger=\frac{\D}{\D x}+\phi(x)$. $H\psi=E\psi$ with $E=k^2$ can be cast in
the form
\begin{eqnarray}
   Q         \psi &=& k \chi \\
  \label{eq5}
   Q^\dagger \chi &=& \left(k-\frac1kV(x)\right) \psi
\end{eqnarray}
We introduce phase $\theta$ and envelope $\EXP{\xi}$ variables~: 
\begin{eqnarray}
  \label{phaseform1}
  \psi(x) & = & \phantom{-} \EXP{\xi(x)}\sin\theta(x) \\
  \label{phaseform2}
  \chi(x) & = & -           \EXP{\xi(x)}\cos\theta(x)
\end{eqnarray}
with initial conditions $\theta(0)=0$ and $\xi(0)=0$. The phase is related to
the Ricatti variable by $z=-\frac{Q\psi}{\psi}=k\cotg\theta$. The interest to
deal with this couple of variables lies in the basic idea of the phase
formalism, {\it i.e.}
the node counting method~: the IDoS coincides with the number of nodes of the
wave function that can be obtained from the evolution of the cumulative phase.
The Lyapunov exponent (inverse localization length) is defined as the rate of
increase of the logarithm of envelope. Therefore
$N(E)=\lim_{x\to\infty}\frac{\theta(x)}{x\pi}$ and
$\gamma(E)=\lim_{x\to\infty}\frac{\xi(x)}{x}$, where we have omitted average
thanks to self-averaging. These expressions give the most simple way to obtain
spectrum and localization length from a practical point of view (for numerical
calculations).

Phase and envelope obey the differential equations~:
\begin{eqnarray}
  \label{pf1}
  \frac{\D\theta}{\D x} &=& 
        k - \frac{V(x)}k \sin^2\theta + \phi(x)\,\sin2\theta \\
  \label{pf2}
  \frac{\D\xi}{\D x} &=& 
            \frac{V(x)}{2k} \sin2\theta - \phi(x)\,\cos2\theta
\end{eqnarray}

\vspace{0.25cm}

\noindent{\bf Negative part of the spectrum~:}
\mathversion{bold}$E=-k^2$.--\mathversion{normal} If we perform the same
manipulations with $E=-k^2$, we obtain~:
\begin{eqnarray}
  \frac{\D\theta}{\D x} &=& 
        k\cos2\theta - \frac{V(x)}k \sin^2\theta + \phi(x)\,\sin2\theta \\
  \frac{\D\xi}{\D x} &=& k\sin2\theta +
            \frac{V(x)}{2k} \sin2\theta - \phi(x)\,\cos2\theta
\end{eqnarray}

\vspace{0.25cm}

\noindent{\bf Invariant measure for the phase}.--
Using (\ref{theorem}) we can write (for a positive energy)~:
\begin{equation}
  \label{SDETheta}
  \D\theta \eqlaw
        k\,\D x + \tilde\beta(\theta)\, \D W(x)\hspace{0.25cm}\mbox{(Stratonovich)}
\end{equation}
where $\tilde\beta(\theta)=\sqrt{\frac{\sigma}{k^2} \sin^4\theta+g\sin^22\theta}$.
The related FPE reads
$\partial_xP(\theta;x)=F_\theta{}P(\theta;x)=$ where
$F_\theta=-k\partial_\theta+\frac12[\partial_\theta\tilde\beta(\theta)]^2$ is the
forward generator. 
The current of  the phase through the interval $[0,\pi]$ is the number of
zeros of the wave function per unit length $N(E)$. 
The stationary solution for constant current
$N(E)=[k-\frac12\tilde\beta(\theta)\partial_\theta\tilde\beta(\theta)]P(\theta)$ is~:
\begin{equation}
  \label{disTheta}
  P(\theta) = \frac{2N(E)}{\tilde\beta(\theta)}
  \int_\theta^\pi\frac{\D\theta'}{\tilde\beta(\theta')}
  \EXP{ 2k\int_{\theta'}^\theta\frac{\D\theta''}{\tilde\beta(\theta'')^2} }
\end{equation}
the IDoS is given by normalizing the distribution.

\subsection{From multiplicative to additive noise}

We have obtained the expression of the IDoS, which is given by
normalizing the distribution (\ref{disRicatti}) or the distribution
(\ref{disTheta}) and is expressed as a double integral. The analysis
of the random process and of its distribution is however made more
simple by converting the SDE for the Ricatti variable (\ref{SDEZ}) or
the phase (\ref{SDETheta}), that include multiplicative noises, into a
SDE with additive noise. For that purpose we perform the following
change of variable~:
\begin{equation}
  z \eqdef \sqrt{\frac{\sigma}{4g}}\,\sinh\varphi
\end{equation}
(that maps \RR\ to \RR). The relation with the phase variable is
$\cotg\theta=\sqrt{\frac{\sigma}{4g|E|}}\,\sinh\varphi$. The new variable obeys
the SDE
\begin{equation}
  \label{SDEVarphi}
  \D\varphi = -\sqrt{\frac{\sigma}{4g}}
  \left[
    \cosh\varphi + 
    \left( \frac{4gE}{\sigma} -1  \right)\frac1{\cosh\varphi}
  \right]\,\D x
  + \sqrt{4g}\,\D W(x)
  =-U'(\varphi)\,\D x+\sqrt{4g}\,\D W(x)
\end{equation}
where we introduced the potential~:
\begin{equation}
  \boxed{
  U(\varphi) = \sqrt{\frac{\sigma}{4g}}
   \left[
    \sinh\varphi + 
    \left( \frac{4gE}{\sigma} -1  \right)\arctan(\sinh\varphi)
  \right] 
  }
\end{equation}
Note that $U(\varphi)=\mathcal{U}(z=\sqrt{\frac{\sigma}{4g}}\,\sinh\varphi)$.

\begin{figure}[!ht]
  \centering
  \includegraphics[scale=0.4]{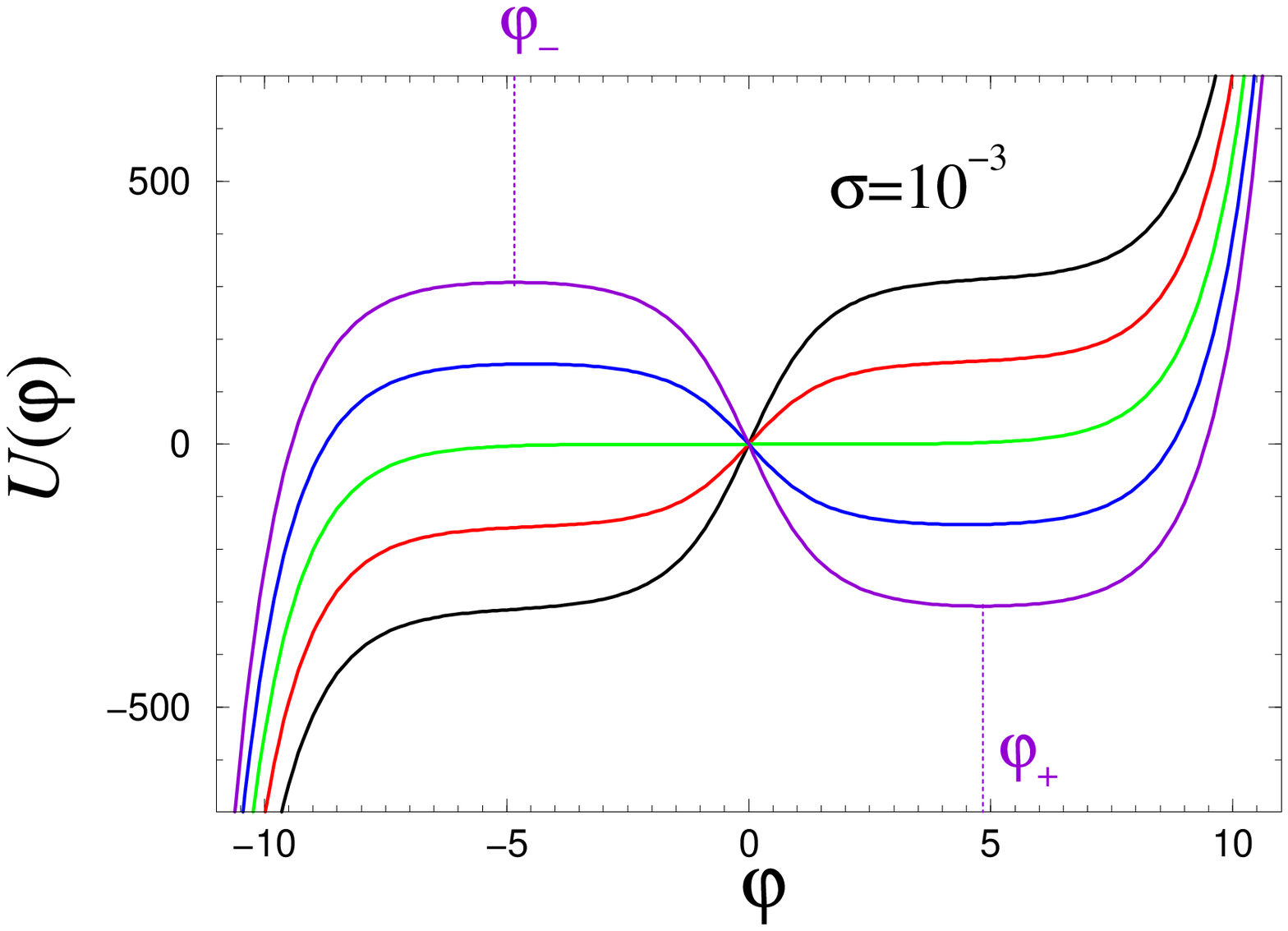}
  \hspace{0.5cm}
  \includegraphics[scale=0.4]{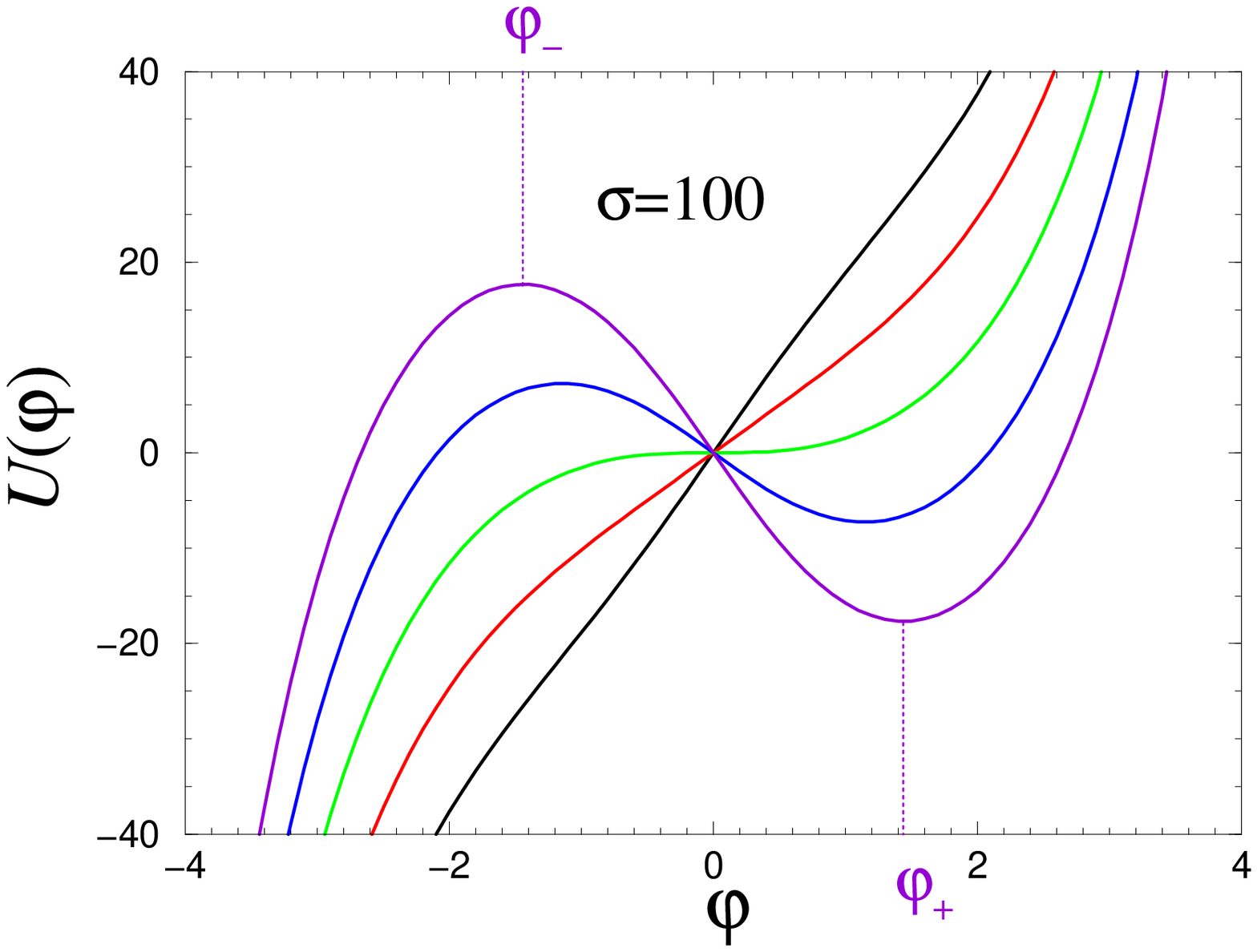}
  \caption{\it Potential $U(\varphi)$ for $g=1$.
    {\sf Left~:} $\sigma=0.001$ with energy $E=10$, $5$, $0$, $-5$, $-10$.
    {\sf Right~:} $\sigma=100$ with energy $E=100$, $50$, $0$, $-50$, $-100$.
     }
  \label{fig:potential}
\end{figure}

In order to get the IDoS we contruct the stationary solution of the FPE
$\partial_x\mathcal{P}(\varphi;x)=F_\varphi\mathcal{P}(\varphi;x)$ where
$F_\varphi=\partial_\varphi{}U'(\varphi)+2g\partial_\varphi^2$ is the forward
generator. The stationary solution for a constant current $-N(E)$ (the variable
$\varphi$ goes from $+\infty$ to $-\infty$, therefore currents for the phase
$\theta$ and for $\varphi$ are opposite) reads~:
\begin{equation}
  \label{disVarphi}
  \mathcal{P}(\varphi) = \frac{N(E)}{2g}\,\EXP{-\frac1{2g}U(\varphi)}
  \int_{-\infty}^\varphi\D\varphi'\,\EXP{\frac1{2g}U(\varphi')}
\end{equation}
which can also be directly obtained from (\ref{disRicatti}) or (\ref{disTheta}) since 
$\D\varphi=-\sqrt{4g}\frac{\D\theta}{\tilde\beta(\theta)}=\sqrt{4g}\frac{\D{}z}{\beta(z)}$.

An alternative way to obtain the IDoS, that will help the discussion and will
be used later, is to introduce the $n$-th moment of the ``time'' $x$ needed by
the process $\varphi(x)$ to reach $-\infty$, starting from
$\varphi(0)=\varphi$ (spatial coordinate $x$ plays the role of the ``time''
and variable $\varphi$ of the position). This problem is a first exit
problem~\cite{Gar89}. The moments are given by solving the equation
$B_\varphi{}T_n(\varphi)=-nT_{n-1}(\varphi)$ where
$B_\varphi=-U'(\varphi)\partial_\varphi+2g\partial_\varphi^2$ is the backward
Fokker-Panck generator. The solution is constructed for absorbing boundary
condition at $-\infty$ and reflecting boundary at $+\infty$~: $T_n(-\infty)=0$
and $\partial_\varphi{}T_n(+\infty)=0$. We find (see Ref.~\cite{Gar89} or
appendix of Ref.~\cite{Tex00})
\begin{equation}
  \label{momentsTn}
  T_n(\varphi) = \frac{n}{2g}
  \int_{-\infty}^\varphi\D\varphi'\,\EXP{\frac1{2g}U(\varphi')}
  \int_{\varphi'}^{+\infty}\D\varphi''\,\EXP{-\frac1{2g}U(\varphi'')}\,
  T_{n-1}(\varphi'')
\end{equation}
$T_n(+\infty)$ corresponds to the $n$-th moment of the time needed by random
process $\varphi$ to cross $\RR$, therefore the moment of the distance $\ell$
between two consecutive nodes of the wave function $\psi(x;E)$. Let us
emphasize on this 
point. We call $\ell_i$ the distance between the two consecutive nodes of the
wave function~: $\psi(0)=\psi(\ell_1)=\psi(\ell_1+\ell_2)=\cdots=0$. The
problem of first exist problem is defined as $\varphi(x_i)=+\infty$ and
$\varphi(x_i+\ell_i)=-\infty$ with $\varphi(x)$ finite for
$x\in]x_i,x_i+\ell_i[$. The random variable is $\ell_i$ and
$\smean{\ell^n}\equiv{}T_n(+\infty)$. Note that all distances are {\it i.i.d.}
due to the fact that potential have a vanishing correlation length\footnote{
  In general distances $\ell_i$ are decorrelated if correlation length is
  smaller than the length over which deterministic dynamics drives
  $\varphi(x)$ to $\infty$.
}.

The IDoS per unit length is the average number of nodes of $\psi(x;E)$
per unit length, what
corresponds to the inverse average distance between two consecutive
nodes~:
\begin{equation}
  N(E)^{-1} = T_1(+\infty) 
\end{equation}
therefore
\begin{equation}
  \label{IDoS}
 \boxed{
  N(E)^{-1} = \frac{1}{2g}
  \int_{-\infty}^{+\infty}\D\varphi\,\EXP{\frac1{2g}U(\varphi)}
  \int_{\varphi}^{+\infty}\D\varphi'\,\EXP{-\frac1{2g}U(\varphi')}
  }
\end{equation}
that coincides with the normalization of the distribution (\ref{disVarphi}).
We will extract limiting behaviours of this exact expression by analyzing more
precisely the dynamics of the random process~$\varphi(x)$.

We first remark that the derivative of the potential at the origin is
\begin{equation}
  U'(0) = \sqrt{\frac{4g}{\sigma}} E
\end{equation}
For $E>0$ the potential is monotonous. 

For $E<0$ it developes a local minimum able to trap the process during a
finite ``time''. In this latter case the local minimum of the potential is at
$\varphi_+>0$ and the top of the barrier at $\varphi_-=-\varphi_+$~:
\begin{equation}
  \label{extrema}
  \sinh\varphi_\pm = \pm\sqrt{\frac{4g|E|}{\sigma}}
\end{equation}
We easily check that $U''(\varphi_\pm)=\pm2\sqrt{-E}$ for $E<0$.

\vspace{0.25cm}

\noindent{\bf Important energy scales.--}
We will identify later the relevant energy scales in the problem. In each
reagime ($g^3\ll\sigma$ or $g^3\gg\sigma$) two energy scales matter~: the two
largest scales among $\sigma/g$, $\sigma^{2/3}$, $\sqrt{g\sigma}$ and $g^2$.
\begin{itemize}
\item For small supersymmetric noise $g^3\ll\sigma$, the two relevant energy
  scales are $\sigma^{2/3}$ and $\sigma/g$.
\item For large supersymmetric noise $g^3\gg\sigma$, the two energy scales are
  $\sqrt{g\sigma}$ and $g^2$.
\end{itemize}

\mathversion{bold}
\subsection{Density of states for positive energies for ${g}^3\gg\sigma$}
\mathversion{normal}

The supersymmetric Hamiltonian is characterized by a purely positive spectrum
(which follows from the structure $H_\mathrm{susy}=Q^\dagger{}Q$) which
presents the famous Dyson singularity at zero
energy~\cite{OvcEri77,BouComGeoLeD90}~:
\begin{equation}
  N^{(\sigma=0)}(E)\sim \frac{g}{\ln^2(g^2/E)}
  \hspace{0.25cm}\mbox{for }
  E\to0
\end{equation}
therefore it vanishes at zero energy~: $N^{(\sigma=0)}(E=0)=0$. What is the
fraction of states that migrate to $\RR^-$ when a very small white noise
$V(x)$ breaking the supersymmetry is added to~$H_\mathrm{susy}$~?

\begin{figure}[!ht]
  \centering
  \begin{tabular}{cc}
  \includegraphics[width=0.45\textwidth]{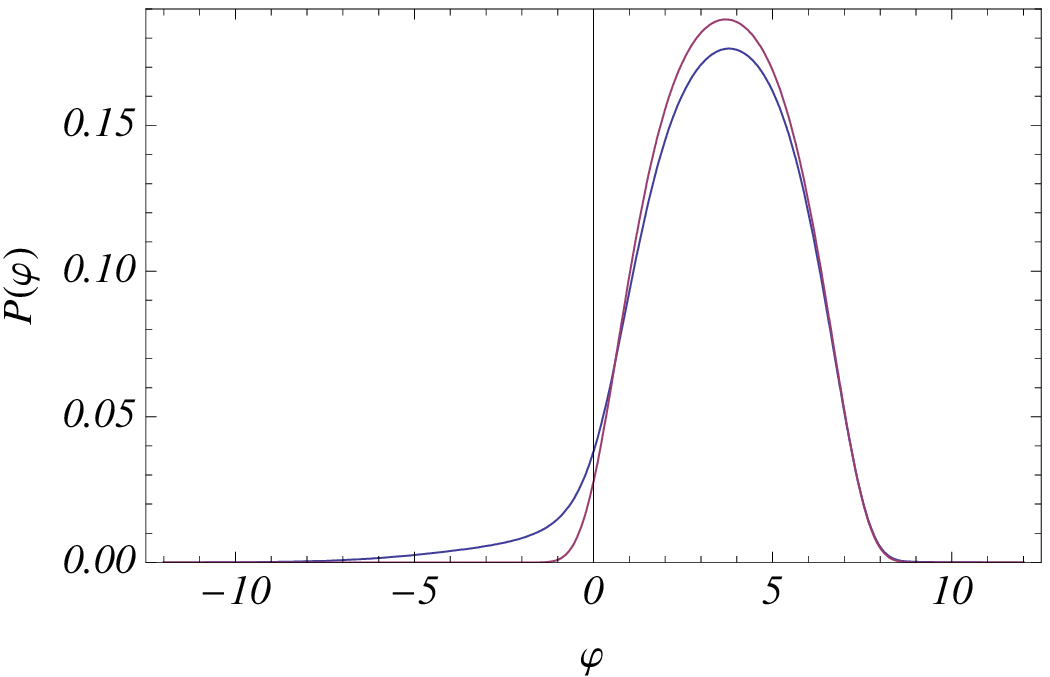}
  &
  \includegraphics[width=0.45\textwidth]{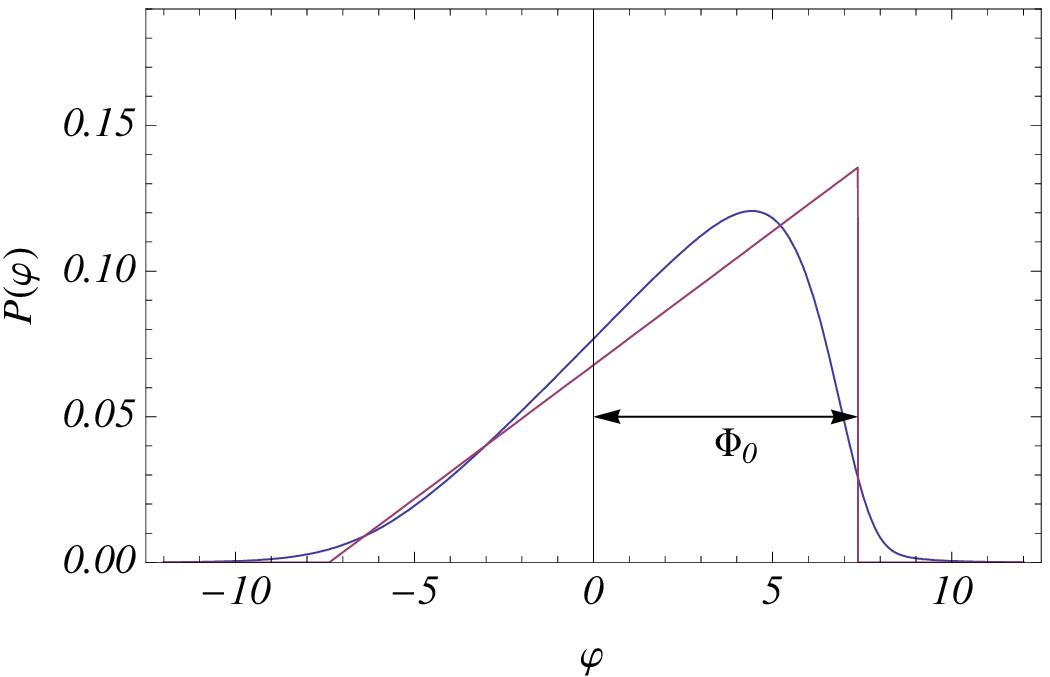}
  \\
  (a) & (b)\\
  {} & {}\\
\  \includegraphics[width=0.45\textwidth]{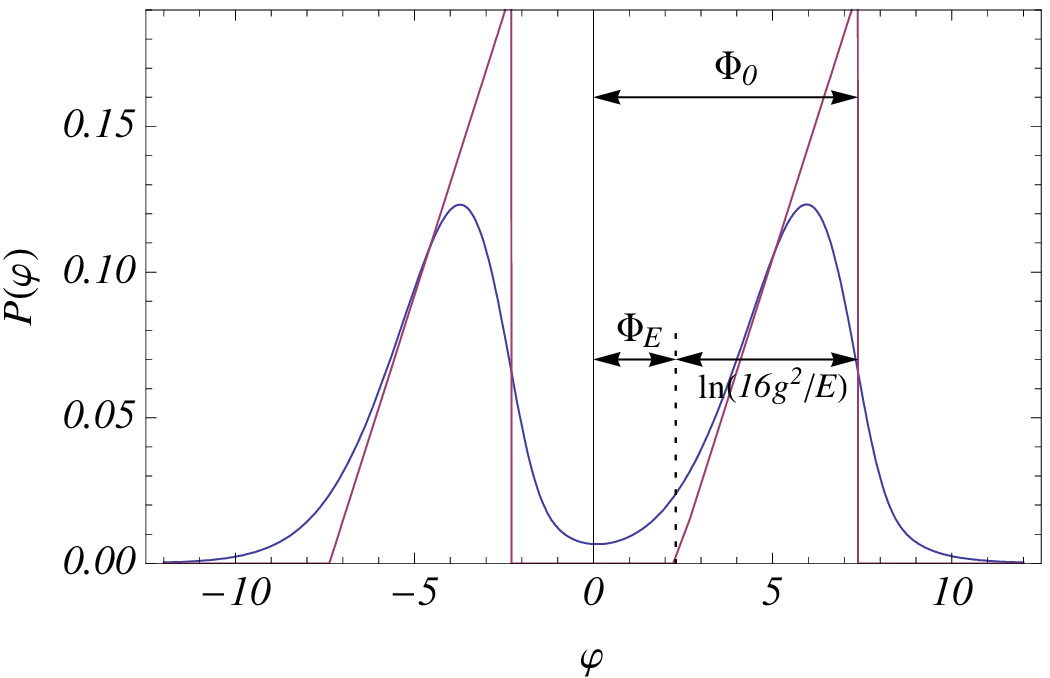}
  &
  \includegraphics[width=0.45\textwidth]{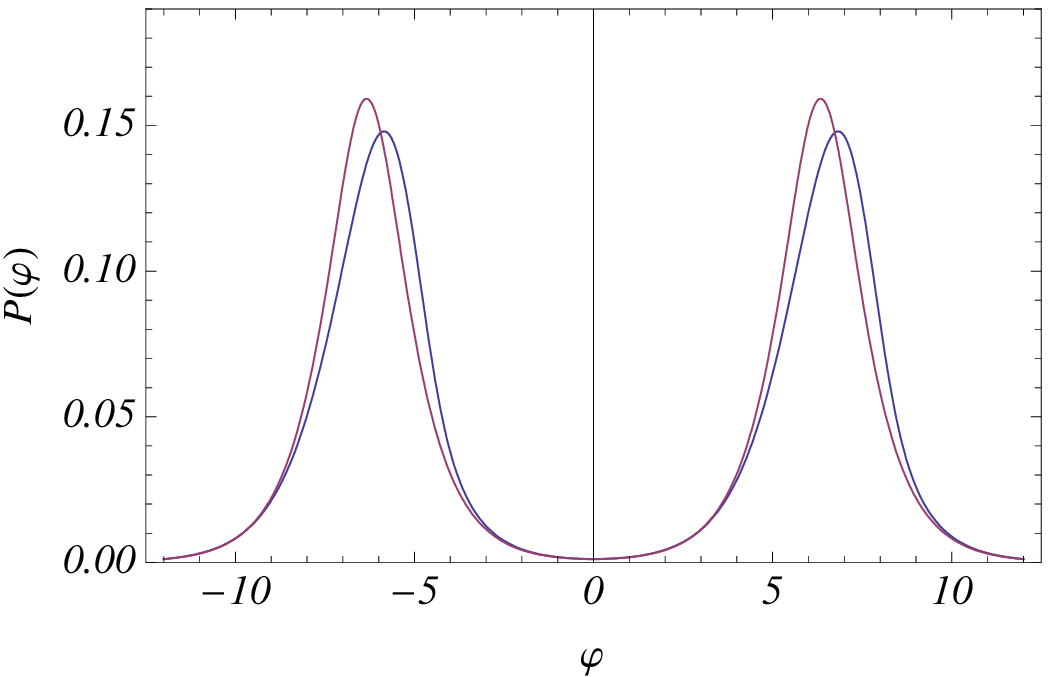}\\
  (c) & (d)
  \end{tabular}
  \caption{\it Distribution $\mathcal{P}(\varphi)$ (blue curves)
    for (a) $E=-0.01$, (b) $E=0$, (c) $E=0.1$ and (d) $E=2$ (with
    $\sigma=10^{-4}$ \& $g=1$). These behaviours can be understood from the
    shapes of the potential of the left part of figure~\ref{fig:potential}.
    For $E=-\sqrt{\sigma{}g}=-0.01$~: the distribution is compared to
    $\exp[-\frac1g\sqrt{|E|}\cosh(\varphi-\varphi_+)]$.
    For the other energies, the distribution is compared to approximations
    obtained in the text.}
  \label{fig:glups}
\end{figure}

\vspace{0.25cm}

\noindent{\bf Band center~: }
\mathversion{bold}$|E|\ll\sqrt{g\sigma}$\mathversion{normal}.-- In the SDE
(\ref{SDEVarphi}), the exponential nature of the potential allows a decoupling
of the deterministic force and the Langevin force (this works for
${g}^3\gg\sigma$ only). We introduce the value for which the two forces are of
the same order~: $|U'(\Phi_0)|\eqdef4g$~:
\begin{equation}
  \Phi_0\simeq\ln\left(16\sqrt{g^3/\sigma}\right)
\end{equation}
In the interval $[\Phi_0,+\infty[$, the dynamics of the random process is
governed by the deterministic force. The process, starting from $+\infty$,
reaches $\Phi_0$ very fast. Then its dynamic is governed by the Langevin force
in $[-\Phi_0,\Phi_0]$. Upon arrival at $-\Phi_0$ it is driven very fast to
$-\infty$ by the deterministic force. This allows to map the problem to the
problem of free diffusion $\D\varphi\simeq\sqrt{4g}\,\D{W}(x)$ on the interval
$[-\Phi_0,\Phi_0]$, with reflecting boundary condition at $+\Phi_0$ and
absorbing boundary condition at $-\Phi_0$ (a similar approximation was used in
\cite{Tex00} to study the supersymmetric Hamiltonian at finite energy $E\ll{g^2}$). We immediatly conclude that the
average ``time'' is
$N(0)^{-1}=T_1(+\infty)\simeq\frac{(\mathrm{distance})^2}{\mathrm{diffusion}}=\frac1{4g}(2\Phi_0)^2$.
Therefore a fraction of states
\begin{equation}
  N(0)\sim  \frac{g}{\ln^2(g^3/\sigma)}
\end{equation}
have migrated to~$\RR^-$.

Let us analyze the structure of the distribution $\mathcal{P}(\varphi)$, given
by (\ref{disVarphi}), in the low energy limit. For $\varphi\lesssim-\Phi_0$ we
have $|U'(\varphi)|\gg2g$ therefore
$\EXP{U(\varphi)/2g}$ is extremely small and the integral over $\varphi'$ is
dominated by the close neighbourhood of $\varphi$~:
\begin{equation}
  \label{onlydrift}
  \mathcal{P}(\varphi) \simeq \frac{N(E)}{|U'(\varphi)|} 
  \simeq
  \frac{N(E)}{4g}\,\frac{\cosh\Phi_0}{\cosh\varphi} \sim \EXP{\varphi+\Phi_0}
  \hspace{0.5cm}\mbox{for }\varphi\lesssim-\Phi_0
\end{equation}
This approximation reflects the fact that when deterministic evolution dominates
$\mathrm{Velocity}(\varphi)=\frac{\D\varphi}{\D{}x}\simeq-U'(\varphi)$ the
distribution is $\mathcal{P}(\varphi)\propto1/|\mathrm{Velocity}(\varphi)|$.

In the intermediate interval $[-\Phi_0,\Phi_0]$, $\EXP{U(\varphi)/2g}$ is
almost flat and the distribution is linear in this interval
\begin{equation}
  \mathcal{P}(\varphi) \simeq 
  \frac{N(E)}{2g}\,
  \left[
      \frac12\EXP{[U(-\Phi_0)-U(\varphi)]/2g}
    + (\varphi + \Phi_0)
  \right]
  \sim\varphi + \Phi_0
  \hspace{0.5cm}\mbox{for }-\Phi_0\lesssim\varphi\lesssim+\Phi_0
\end{equation}
where the first term is the contribution of the interval $]-\infty,-\Phi_0]$
to the integral (\ref{disVarphi}).
Finally we find for the last interval 
\begin{equation}
    \mathcal{P}(\varphi) \simeq 
  \frac{N(E)}{2g}\,
  \left[
      \frac12\EXP{[U(-\Phi_0)-U(\varphi)]/2g}
    + (\varphi +  \Phi_0)\EXP{[U(\Phi_0)-U(\varphi)]/2g}
    + \frac12 \frac{\cosh\Phi_0}{\cosh\varphi}
  \right]
  \hspace{0.5cm}\mbox{for }\Phi_0\lesssim\varphi
\end{equation}
It decreases exponentially~: $\mathcal{P}(\varphi)\sim\EXP{-\varphi+\Phi_0}$.
The curve is plotted on figure~\ref{fig:glups}. Adding times spent in the
three intervals gives the normalization
$N(0)^{-1}\simeq\frac1{4g}+\frac1g\ln^2(16\sqrt{g^3/\sigma})+\frac1{4g}$~:
\begin{equation}
  \boxed{
  N(E)\simeq \frac{4g}{\ln^2(2^8g^3/\sigma)+2}
  }
  \hspace{0.25cm}\mbox{for }
  |E|\ll\sqrt{g\sigma}
\end{equation}

\vspace{0.25cm}

\noindent{\bf Intermediate energies~:}
\mathversion{bold}$\sqrt{g\sigma}\ll{E}\ll{g}^2$\mathversion{normal}.-- In
this limit the potential developes a double plateaux structure as suggested on
figure~\ref{fig:potential}. Once again we use the fact that the deterministic
force depends exponentially on $\varphi$ to decouple the effects of the
Langevin force and the deterministic force.
The equation $|U'(\varphi)|=4g$ possesses now four solutions~:
$\varphi=\pm\Phi_0$ defined above and  $\varphi=\pm\Phi_E$ with
\begin{equation}
  \Phi_E\simeq\ln\left(E/\sqrt{\sigma g}\right)
\end{equation}
The Langevin force dominates the evolution in intervals corresponding to
plateaux of $U(\varphi)$, of width $\Phi_0-\Phi_E\simeq\ln(16g^2/E)$, while the
deterministic force governs the evolution on the other intervals. Let us
follow the evolution of the process $\varphi(x)$.
({\it i})~In the interval $[\Phi_0,\infty[$ the deterministic force,
$\D\varphi\simeq-U'(\varphi)\D{}x\simeq-\sqrt{\frac{\sigma}{4g}}\cosh\varphi\,\D{}x$,
drives the process from $\varphi=\infty$ to $\varphi=\Phi_0$ in a ``time'' $1/(4g)$.
({\it ii})~In $[\Phi_E,\Phi_0]$ the Langevin force dominates~:
$\D\varphi\simeq\sqrt{4g}\,\D{}W(x)$. Given that one is reflected at $\Phi_0$,
the average ``time'' required to reach $\Phi_E$ for the first time is
$\frac{(\mathrm{distance})^2}{\mathrm{diffusion}}=\frac1{4g}(\Phi_0-\Phi_E)^2$.
({\it iii})~In $[-\Phi_E,\Phi_E]$ the deterministic force dominates
$\D\varphi\simeq-U'(\varphi)\D{}x\simeq-\sqrt{\frac{4g}{\sigma}}E\frac1{\cosh\varphi}\,\D{}x$
and drives the process from one edge of the interval to the other in a ``time''
$1/(2g)$.
({\it iv})~In $[-\Phi_0,-\Phi_E]$, the Langevin force dominates~: the process
crosses the interval in an average ``time'' $\frac1{4g}(\Phi_0-\Phi_E)^2$.
({\it v})~Finally the deterministic force brings the process from $-\Phi_0$
to $-\infty$ in a ``time'' $1/(4g)$.

The analysis of the distribution (\ref{disVarphi}) follows the same logic. In
the two intervals where motion is diffusive (where the process spends most of
the time)~:
\begin{equation}
   \mathcal{P}(\varphi) \simeq 
  \frac{N(E)}{2g}\,
  \left\{
  \begin{array}{llr}
    \varphi +  \Phi_0  & \mbox{ for } & -\Phi_0\lesssim\varphi\lesssim-\Phi_E \\[0.2cm]
    \varphi -  \Phi_E  & \mbox{ for } & \Phi_E\lesssim\varphi\lesssim\Phi_0
  \end{array}
  \right.
\end{equation}
(see figure~\ref{fig:glups}). Normalizing this distribution gives
$N(E)^{-1}\simeq\frac1{g}+2\frac{(\Phi_0-\Phi_E)^2}{4g}\simeq\frac1{2g}\ln^2(16g^2/E)$
therefore we recover the usual Dyson singularity (the scalar potential $V(x)$
plays no role)~:
\begin{equation}
  \boxed{
  N(E)\simeq \frac{2g}{\ln^2(16g^2/E)+2}
  }
  \hspace{0.25cm}\mbox{for }
  \sqrt{g\sigma}\ll{E}\ll{g}^2
\end{equation}

\vspace{0.25cm}

\noindent{\bf Large energies~: }
\mathversion{bold}${E}\gg{g}^2$\mathversion{normal}.-- Finally, for
completeness, we give the distribution in the high energy limit. In this case
the phase distribution is almost flat $P(\theta)\simeq1/\pi$ therefore the
distribution for $\varphi$ presents the double peak structure~:
\begin{equation}
  \mathcal{P}(\varphi)\simeq\frac1\pi
  \frac{\sinh\varphi_+\cosh\varphi}{ \sinh^2\varphi_+ + \sinh^2\varphi}
\end{equation}
where $\varphi_+$ is defined by (\ref{extrema}). The two peaks are associated
with inflection points of the potential $U(\varphi)$ where the force
is minimum (note that
$\pm\tilde\varphi_+\simeq\varphi_\pm$). The IDoS is given by the free IDoS
$N(E)\simeq\frac1\pi\sqrt{E}$.

\subsection{Lifshits tail \label{sec:Lifshits}}

In this paragraph we analyze the tail of the IDoS in the region of
rarefaction of states, that is for $E\to-\infty$.

For negative energies, the process $\varphi(x)$ is trapped by the well at
$\varphi=\varphi_+$ a very long ``time'' where positions $\varphi_\pm$ of the
extrema of the potential are given by (\ref{extrema}). The average ``time''
needed to exit the well due to a fluctuation (Langevin force) is given by
the Arrhenius formula. The height of the potential barrier is given by~:
\begin{equation}
  \frac1{2g}[U(\varphi_-)-U(\varphi_+)]
  =\sqrt{\frac{\sigma}{4g^3}}\,
   F\left(\frac{4g|E|}{\sigma}\right)
\end{equation}
with 
\begin{align}
  F(x) &\eqdef 2(x+1)[\arctan(\sqrt{x+1}+\sqrt{x})-\pi/4] - \sqrt{x}\nonumber\\
  &=
  \left\{
    \begin{array}{ll}
      \frac23 x^{3/2} + O(x^{5/2}),      & \text{for } x\ll 1              \\[0.2cm]
      \frac\pi2 x     -2\sqrt{x}+\frac\pi2+O(x^{-1/2}),      & \text{for } x\gg 1   
    \end{array}
  \right.
\end{align}
Assuming $\frac1{2g}[U(\varphi_-)-U(\varphi_+)]\gg1$ we can expand integrands in
(\ref{IDoS}). As we can see on figure ~\ref{fig:potential}, in the limit
$g^3\ll\sigma$ the potential $U(\varphi)$ is parabolic near its extrema and
we can use formula (A22) of Ref.~\cite{Tex00}. However in the regime
$g^3\gg\sigma$ the parabolic approximation is not correct. In
this latter case, noting that 
\begin{equation}
  U(\varphi)\underset{\varphi \sim \varphi_\pm}{\simeq}
  \mathrm{const.} \pm 2\sqrt{|E|}\,\cosh(\varphi-\varphi_\pm)
  \:,
\end{equation}
we obtain
\begin{equation}
  \label{generalLif}
  \boxed{
  N(E) \simeq 
  \frac{g}{2}
  \left[\EXP{\frac{\sqrt{|E|}}{g}}K_0\left(\frac{\sqrt{|E|}}{g}\right)\right]^{-2}
  \exp-\sqrt{\frac{\sigma}{4g^3}} F\left(\frac{4g|E|}{\sigma} \right)
  }
 \hspace{0.5cm}
  \mbox{for }
 |E|\gg\max{\sigma^{2/3}}{\sqrt{g\sigma}} 
\end{equation}
where $K_0(z)$ is the MacDonald function (modified Bessel function of third kind). 
We can now consider two situations, depending on which among the supersymmetric
noise $\phi(x)$ or the scalar noise $V(x)$ dominates.

\vspace{0.25cm}

\noindent{\bf Small supersymmetric noise}
\mathversion{bold}${g}^3\ll\sigma$\mathversion{normal}.-- In the intermediate
range we recover from (\ref{generalLif}) the Lifshits tail of the Hamiltonian
$H_\mathrm{scalar}=-\frac{\D^2}{\D{}x^2}+V(x)$~\cite{Hal65,LifGrePas88,ItzDro89}
\begin{equation}
  N(E) \simeq \frac{\sqrt{|E|}}{\pi} \exp -\frac{8|E|^{3/2}}{3\sigma}
    \hspace{0.25cm}\mbox{for } \sigma^{2/3} \ll |E| \ll \sigma/g
\end{equation}
The supersymmetric noise does not affect the DoS in this regime.

For larger values of $|E|$ the tail takes the form
\begin{equation}
   \label{IDoSlsn0}
  N(E) \simeq \frac{\sqrt{|E|}}{\pi} 
  \exp\left[ 
     -\frac{\pi|E|}{\sqrt{g\sigma}}
     + 2\frac{\sqrt{|E|}}{g} 
     -\frac\pi4 \sqrt{\frac{\sigma}{g^3}}
      \right]
    \hspace{0.25cm}\mbox{for } |E| \gg \sigma/g
\end{equation}
Eventhough the supersymmetric noise is much smaller than $V(x)$, the behaviour
at largest values of $|E|$ is due to a competition between $\phi$ and~$V$.

\vspace{0.25cm}

\noindent{\bf Large supersymmetric noise}
\mathversion{bold}${g}^3\gg\sigma$\mathversion{normal}.-- Expanding
(\ref{generalLif}), we see that the IDoS presents the limiting behaviours
\begin{equation}
   \label{IDoSlsn1}
   N(E) \simeq \frac{2g}{\ln^2(g^2/|E|)}\, 
  \exp-\frac{\pi|E|}{\sqrt{g\sigma}}
    \hspace{0.25cm}\mbox{for } \sqrt{g\sigma}\ll |E| \ll g^2
\end{equation}
and
\begin{equation}
   \label{IDoSlsn2}
   N(E) \simeq \frac{\sqrt{|E|}}{\pi} 
  \exp\left[ 
     -\frac{\pi|E|}{\sqrt{g\sigma}}
     + 2\frac{\sqrt{|E|}}{g} 
      \right]
    \hspace{0.25cm}\mbox{for } |E| \gg g^2
\end{equation}
It is interesting to note that the prefactors coincide with the limiting
behaviours obtained for positive energies~:
$N(E)\simeq\frac{2g}{\ln^2(g^2/E)}$ for $\sqrt{g\sigma}\ll{E}\ll{}g^2$ and
$N(E)\simeq\frac1{\pi}\sqrt{E}$ for $E\gg{}g^2$.


\subsection{Extreme value spectral statistics\label{sec:extreme}}

Up to now we have studied spectral properties through the density of states.
In this section we consider another property of the spectrum~: the problem of
extreme value statistics for the eigenvalues of the Hamiltonian
(\ref{eqn:hamiltonian}). Let us formulate the problem~: for a given
realization of the potential, the spectral (Sturm-Liouville) problem
$H\psi(x)=E\psi(x)$ for boundary conditions $\psi(0)=\psi(L)=0$ has a discrete
set of solutions $\mathrm{Spec}(H)=\{E_n\}$ (we assume that label corresponds
to rank the eigenvalues as $E_1<E_2<E_3<\cdots$). We ask the question~: what
is the distribution 
\begin{equation}
  \label{Wn}
  W_n(E) = \smean{ \delta(E - E_n) }
\end{equation}
of the $n-$th eigenvalue~? These distributions give a much more precise
information on the spectrum than the density of states, what is already clear
from the relation $\sum_{n=1}^\infty{}W_n(E)=L\rho_L(E)$ where $\rho_L(E)$ is
the average DoS per unit length accounting for the Dirichlet boundary
conditions at $x=0$ and $x=L$ (when $L\to\infty$ the sensitivity to the
boundary conditions disappears~: $\lim_{L\to\infty}\rho_L(E)=N'(E)$, where
$N(E)$ is the IDoS per unit length of the infinite system studied above). The
distribution $W_n(E)$ gives the probability to find the $n$-th eigenvalue at
$E$ whereas the DoS $\rho_L(E)$ tells us the probability to find {\it any}
eigenvalue at~$E$.

The study of extreme value statistics in various contexts has attracted a lot of
attention. Extreme value statistics of uncorrelated and identically
distributed variables were classified long time ago (Gumbel for an
exponentially decreasing distribution, Fr\'echet for a power law \& Weibull
for distribution with bounded support \cite{Gum54,Gum58}). Extreme value
statistics for correlated variables is a much more difficult task. A famous
example is the Tracy-Widom distribution for eigenvalues of Gaussian random
matrices \cite{TraWid93,TraWid02}. There has been a renewed interest in such
problems in the last years (see for example Refs.~\cite{MajKra03,ComLebMaj07}).

The question of extreme value statistics of a 1d random Hamiltonian
was first addressed in Ref.~\cite{GreMolSud83} for the Hamiltonian
$H=-\frac{\D^2}{\D{}x^2}+\sum_nv_n\delta(x-x_n)$ where positions are
uncorrelated and uniformly distributed~; weights $v_n$ are positive,
uncorrelated and distributed according to a Poisson law. The case of
the Hamiltonian $H_\mathrm{scalar}=-\frac{\D^2}{\D{}x^2}+V(x)$ where
$V(x)$ is a white noise was studied in Ref.~\cite{McK94} where
$W_1(E)$ was derived. This result was generalized in Ref.~\cite{Tex00}
where it was shown that the distributions $W_n(E)$ are Gumbel laws
when $L\to\infty$~: despite eigenvalues $E_n$ are random variables
{\it a priori} correlated, extreme value distributions coincide with
those of {\it uncorrelated} variables.  Such an absence of spectral
correlations is a consequence of the strong localization of the wave functions
in this regime~\cite{Mol81}.  Extreme value spectral statistics for
the supersymmetric Hamiltonian (\ref{Hsusy}) near the delocalization
transition was also considered in Ref.~\cite{Tex00}~; it was noticed
that in this case the distributions $W_n(E)$ do not coincide with extreme
value statistics for uncorrelated variables, a consequence of spectral
correlations near the delocalization transition.

We first consider the limit of strong supersymmetric disorder
$g^3\gg\sigma$.  When a small scalar noise is added to the
supersymmetric Hamiltonian, we have seen that the spectrum is not
anymore constrained to be in $\RR^+$. A simple way to obtain the
typical ground state
energy is to write that $LN(E_1)\sim1$ from which we obtain
$E_1\sim-\sqrt{\sigma{}g}\,\ln{}L$.  Since the exponential tail of the
IDoS is usually associated to strongly localized states, what will be
supported by the study of localization in section~\ref{sec:loc}, we expect
that the distributions (\ref{Wn}) are similar to the one obtained for
$H_\mathrm{scalar}$ (Gumbel laws). 
This is the aim of
the following paragraph to show this statement explicitely.

We first assume that the length of the system is sufficiently long so that the
support of (\ref{Wn}) is in $\RR^-$ with energies far from the band center
$|E|\gg\sqrt{g\sigma}$. In this case, the process $\varphi(x)$ is trapped by
the well of the potential $U(\varphi)$. The ``time'' $\ell$ needed by the
process to go from $+\infty$ to $-\infty$ ($\ell$ is the distance between two
consecutive nodes of the wave function) is dominated by the time needed to
exit the well. Its moments~(\ref{momentsTn}) are given by
$\smean{\ell^n}=T_n(+\infty)\simeq{n!}[T_1(+\infty)]^n$ what corresponds to a
Poisson law. As a consequence it was shown in Ref.~\cite{Tex00} that
\begin{equation}
  \label{ResTex00}
  W_n(E) \simeq  L\rho(E)\,\frac{\left[LN(E)\right]^{n-1}}{(n-1)!}\,\EXP{-L\,N(E)}
\end{equation}
where the IDoS per unit length of the infinite system is given by 
(\ref{IDoSlsn1},\ref{IDoSlsn2}). The
question of which, among (\ref{IDoSlsn1}) or (\ref{IDoSlsn2}), is the behaviour
to be considered in order to analyze $W_n(E)$ depends on where the support
of the distribution is. {\it A priori} for the longest size $L\to\infty$ we
expect that $W_n(E)$ has its support for energies below $-g^2$ whereas for
intermediate length $L$ the support is between $-g^2$ and $-\sqrt{\sigma{}g}$.
This question will be rediscussed more precisely below.

\begin{figure}[htpb]
  \centering
  \includegraphics[scale=0.75]{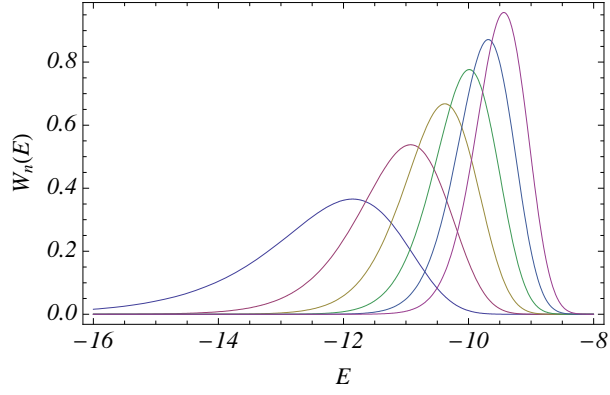}
  \caption{\it nth energy-level distribution $W_n(E)$ for $1\leq{n}\leq6$ with
    $g=10$, $\sigma=1$ and $L=10^4$.
          We see from eqs.~(\ref{EnTyp},\ref{deltaEn}) 
          that the dependence in
          $n$ of the typical energy level and the width are
          $E_n^\mathrm{typ}\simeq{}E_1^\mathrm{typ}+\frac1\pi\sqrt{g\sigma}\ln(n)$ 
          and $\delta{E}_n\simeq\delta{E}_1/\sqrt{n}$, respectively.}
  \label{fig:lowendistr}
\end{figure}

We see from eqs.~(\ref{IDoSlsn1},\ref{IDoSlsn2}) that the density of
states per unit length is well approximated
by~$\rho(E)\simeq\frac{\pi}{\sqrt{g\sigma}}N(E)$. In a first time we
assume that $L$ is sufficiently large so that the support of $W_n(E)$
is below $-g^2$. We can use (\ref{IDoSlsn2}) from which we write
\begin{equation}
  W_n(E) \simeq \frac1{(n-1)!}
  \frac{\pi}{\sqrt{g\sigma}}\left(\frac{L}{\pi}\right)^{n}\:\EXP{-f(E)}
\end{equation}
with
\begin{equation}
 f(E)=-\frac{n}{2}\ln|E|+\frac{n\pi|E|}{\sqrt{g\sigma}}-\frac{2n\sqrt{|E|}}{g}
  + n \frac\pi4\sqrt{\frac{\sigma}{g^3}}
  +\frac{\sqrt{|E|}L}{\pi}\:
  \EXP{ 
    -\frac{\pi|E|}{\sqrt{g\sigma}}
    +\frac{2\sqrt{|E|}}{g}
    -\frac\pi4\sqrt{{\sigma}/{g^3}}
      }
\end{equation}
(note that we have reintroduced the term $\frac\pi4\sqrt{\sigma/g^3}$
neglected in (\ref{IDoSlsn2}) but present in (\ref{IDoSlsn0})~; this
will be useful to discuss the other limit $g^3\ll\sigma$).  It is
convenient to re-scale energy and length as
\begin{equation}
  y =\frac{\pi|E|}{\sqrt{g\sigma}} \qquad\text{and}\qquad 
  \tilde{L}=\frac{(g\sigma)^{1/4}L}{\pi^{3/2}n},
\end{equation}
and furthermore to introduce the quantity
$\varepsilon=(\sigma/\pi^2g^3)^{1/4}$.  Hence, in terms of the new
variables we find $f(E)=g(y)+$const.
for
\begin{equation}
  g(y) = n\, 
  \left[
    -\frac12\ln{y} + y -2\varepsilon\sqrt{y}+ 
    \tilde{\mathcal{L}}\sqrt{y}\,
    \EXP{ -y+2\varepsilon \sqrt{y} } 
  \right]    
  \hspace{0.5cm}\mbox{with}\hspace{0.5cm}
  \tilde{\mathcal{L}}=\tilde{L}\,\EXP{-(\frac{\pi\varepsilon}2)^2}
\end{equation}
The derivative reads
$g'(y) = n\,(1-\frac1{2y}-\frac{\varepsilon}{\sqrt{y}})\,(1-\tilde{\mathcal{L}}\sqrt{y}\,\EXP{-y+2\varepsilon\sqrt{y}})$.
The first paranthesis vanishes for a value of $y$ corresponding to
energy out of the range defined in (\ref{IDoSlsn2})~; it should not be
considered as an extremum.
The extremum $y=\tilde{y}$ is solution of 
\begin{equation}
  \label{ytilde}
  \tilde{\mathcal{L}}
  \sqrt{\tilde{y}}\,\EXP{-\tilde{y}+2\varepsilon\sqrt{\tilde{y}}}=1
\end{equation}
In the limit $L\to\infty$ we find~:
\begin{equation}
  \label{ytilde2}
  \tilde{y} = \ln\tilde{L} + \frac12\ln\ln\tilde{L} 
  + 2\varepsilon\sqrt{\ln \tilde{L}}
  + \left(2-\frac{\pi^2}{4}\right) \varepsilon^2
  +O\left(\frac{\ln\ln\tilde{L}}{\sqrt{\ln\tilde{L}}}\,,
          \frac{\varepsilon}{\sqrt{\ln \tilde{L}}}\right)
\end{equation}
We can easily show that higher derivatives are given by~:
$g^{(k)}(\tilde{y})\simeq{}n\,(-1)^{k}$ for $\tilde{L}\to\infty$ and $k>1$.
Typical value of the energies (value that maximizes $W_n(E)$) is
\begin{equation}
  \label{EnTyp}
  \boxed{
   E_n^\mathrm{typ} = 
  -\frac{\sqrt{g\sigma}}{\pi}\ln\left(\tilde{L}\sqrt{\ln\tilde{L}}\right)
  -\frac{2\,\sigma^{3/4}}{\pi^{3/2}g^{1/4}}\sqrt{\ln \tilde{L}}+\cdots 
  }
\end{equation}
The width of the distribution is independent on the length~:
\begin{equation}
  \label{deltaEn}
  \boxed{
  \delta{E}_n \simeq \frac1\pi\sqrt{\frac{\sigma g}{n}}
  }
\end{equation}
Following \cite{Tex00} we can reconstruct $g(y)$ in the neighbourhood of
$\tilde{y}$ by using the derivatives. An alternative formulation is to expand
$g(\tilde{y}-\frac1{\sqrt{n}}X)$, where
$X=\frac1{\delta{E}_n}(E-E_n^\mathrm{typ})=\sqrt{n}(\tilde{y}-y)$. We can
neglect all terms vanishing in the limit $\tilde{y}\to\infty$
({\it i.e.} $\tilde{L}\to\infty$) since $X\sim1$. 
We have 
\begin{align}
\frac1ng\left(\tilde{y}-\frac1{\sqrt{n}}X\right)
&=-\frac12\ln(\tilde{y}-X/\sqrt{n})+\tilde{y}-X/\sqrt{n}-2\varepsilon \sqrt{\tilde{y}-X/\sqrt{n}}\nonumber \\
&\hspace{0.2\textwidth}+\tilde{L}
\sqrt{\tilde{y}-X/\sqrt{n}}\,\EXP{-\tilde{y}+X/\sqrt{n}+2\varepsilon
  \sqrt{\tilde{y}-X/\sqrt{n}}}
\end{align}
Using (\ref{ytilde}), we obtain
\begin{align}
\frac1ng\left(\tilde{y}-\frac1{\sqrt{n}}X\right)
\underset{\tilde y\to\infty}{\simeq}
\mathrm{const.}-\frac{X}{\sqrt{n}}+\EXP{X/\sqrt{n}}.
\end{align}
Therefore we have recovered the Gumbel law
\begin{equation}
  \label{extreme}
  \boxed{
  W_n(E) \underset{L\to\infty}{=} \frac1{\delta{E}_n}\: 
  \omega_n \left( \frac{E-E_n^\mathrm{typ}}{\delta{E}_n} \right)
  \hspace{0.5cm}\mbox{with}\hspace{0.5cm}
  \omega_n(X) = \frac{n^{n-1/2}}{(n-1)!} 
  \exp{ \left(\sqrt{n}\,X-n\EXP{X/\sqrt{n}}\right) }
  }
\end{equation}
The first distributions are plotted on figure~\ref{fig:lowendistr}.

Let us do several remarks
\begin{itemize}

\item The results (\ref{EnTyp},\ref{deltaEn},\ref{extreme}) have been
  obtained using the asymptotic form of the IDoS (\ref{IDoSlsn2}).
  Therefore it was assumed from the outset of the calculation that
  the support of the distribution (\ref{extreme}) is below $-g^2$. The
  condition $-E_n^\mathrm{typ}\gg{}g^2$ can be recast as a
  condition on the lentgh of the system
  \begin{equation}
     \label{CondL1}
     L \gg \frac{n \pi^{3/2}}{(\sigma g)^{1/4}}\,\EXP{\pi\sqrt{g^3/\sigma}}
  \end{equation}

\item If the length of the system does not satisfy the condition
  (\ref{CondL1}), the support of $W_n(E)$ is shifted above to the
  interval between $-g^2$ and $-\sqrt{\sigma g}$. Therefore the above
  calculation should be redone starting from (\ref{IDoSlsn1}). The
  results are almost similar~: the final distribution (\ref{extreme})
  still holds for the same width (\ref{deltaEn}). Only the behaviour of
  the typical energy changes slightly~:
  \begin{equation}
    \label{EnTyp2}
    \boxed{
     E_n^\mathrm{typ} \simeq
    -\frac{\sqrt{g\sigma}}{\pi}
     \ln\left(
       \frac{
        \tilde{L}'
        }{
        \ln^{2}\frac{\pi\sqrt{g^3/\sigma}}{\ln\tilde{L}'}
        }
        \right)
    }
  \end{equation}
  with $\tilde{L}'=\frac{2gL}{n}$.
  This expression holds when the length of the system is such that 
  \begin{equation}
     \label{CondL2}
    \frac1g \ll L \ll \frac1g\,\EXP{\pi\sqrt{g^3/\sigma}}
  \end{equation}
  The crossover (for $L\sim\EXP{\pi\sqrt{g^3/\sigma}}$)
  obviously corresponds to~$E_n^\mathrm{typ}\sim-g^2$.
  
  For  smaller system sizes $L\lesssim1/g$ we expect the disorder to
  have a perturbative effect and consequently the energy level to be close
  to the free levels $E_n\simeq(\pi n/L)^2$, $n\in\NN^*$.
  Figure~\ref{fig:largeggs} summarizes the different
  regimes for the groundstate energy~$E_1$.
  
\item The introduction of the scalar noise has rather strong consequences on
  the distributions $W_n(E)$. (A) For $\sigma=0$ distributions $W_n(E)$ are
  broad distributions (in particular $E_1^\mathrm{typ}\sim{}g^2\EXP{-gL}$ and
  $\smean{E_1}\sim{}g^2\EXP{-(gL)^{1/3}}$) departing from Gumbel
  distributions, a consequence of spectral correlations \cite{Tex00}. (B) For
  $\sigma\neq0$ the distributions are narrow distributions centered on
  $E_n^\mathrm{typ}\sim-\sqrt{g\sigma}\ln{L}$ and coinciding with Gumbel
  distributions, an indication of absence of spectral correlations. We now
  characterize the crossover scale of scalar noise separating the two
  situations (A) \& (B). Let us reason at {\it fixed} $g$ and $L$ (with
  $L\gg1/g$) and introduce an infinitesimal $\sigma$~: we start from the
  situation (A). If $\sigma$ is increased, the length $L$ fulfills the
  condition (\ref{CondL2}) and the ground state energy is given by
  (\ref{EnTyp2}), provided that at least one state is below $-\sqrt{g\sigma}$.
  This last condition reads
  $L\,N(-\sqrt{g\sigma})\sim\frac{gL}{\ln^2(g^3/\sigma)}\gtrsim1$. This leads
  to the crossover value
  \begin{equation}
    \boxed{
    \sigma_c \sim g^3\, \EXP{-\sqrt{gL}}
    }
  \end{equation}
  separating (A) and (B). Below this value ($\sigma\lesssim\sigma_c$)
  the scalar noise can be ignored.
  Another simple way to obtain this scale is to write
  $-E_n^\mathrm{typ}\gtrsim\sqrt{g\sigma}$, where the typical energy
  is given by~(\ref{EnTyp2}). 

\item{\it Small supersymmetric noise.--} Finally we mention the
  results for $\sigma\gg{g}^3$. If the support of $W_n(E)$ is in the
  interval between $-\sigma/g$ and $-\sigma^{2/3}$ the supersymmetric
  noise does not play any role and we recover the results of
  Ref.~\cite{Tex00} obtained for $H_\mathrm{scalar}$~: the form
  (\ref{extreme}) holds for
  \begin{equation}
    \label{UltEnTyp}
    \boxed{
    E_n^\mathrm{typ} \simeq
    -\left(\frac{3\sigma}{8}\ln\tilde{L}''\right)^{2/3}
    \hspace{0.5cm}\mbox{and}\hspace{0.5cm}
    \delta E_n \simeq \frac{\sigma^{2/3}}{2\sqrt{n}}\left(\ln\tilde{L}''\right)^{-1/3}
    }
  \end{equation}
  with $\tilde{L}''=\frac{L\sigma^{1/3}}{2\pi n}$ (this time the width
  of the distributions vanish in the limit $L\to\infty$).  The
  hypothesis made on the position of the support of $W_n(E)$ implies
  that the length satisfies
  \begin{equation}
    \sigma^{-1/3} \ll L 
    \ll n \sigma^{-1/3}\,\EXP{\frac83\sqrt{\frac{\sigma}{g^3}}}
  \end{equation}
  For longer lengths we recover the behaviours
  (\ref{EnTyp},\ref{deltaEn}).
  Now the term $\varepsilon^2$ of (\ref{ytilde2}) neglected above
  is large, what adds a term to (\ref{EnTyp}).
  We can check that at the crossover 
  ($L\sim\EXP{\frac83\sqrt{\sigma/g^3}}$) both (\ref{EnTyp})
  and (\ref{UltEnTyp}) give~$E_n^\mathrm{typ}\sim-\sigma/g$.

  An illustration of the different regimes is given on
  figure~\ref{fig:smallggs}.
   
  \begin{figure}[h]
    \centering
    \begin{pspicture}(13,3.4)
      \psline[arrowsize=5pt]{->}(0,2)(12,2)
      \psline(0,1.9)(0,2.1)
      \psline(2,1.9)(2,2.1)\psline[linestyle=dotted](2,1.9)(2,0)
      \psline(7,1.9)(7,2.1)\psline[linestyle=dotted](7,1.9)(7,0)
      \psline(8.3,1.9)(8.3,2.1)\psline[linestyle=dotted](8.3,1.9)(8.3,0)
      \rput(12.5,2){$L$}
      \rput(2,2.6){$\frac{1}{g}$}
      \rput(7,2.6){$\frac{\EXP{\pi\sqrt{g^3/\sigma}}}{g}$}
      \rput(8.4,2.6){$\frac{\EXP{\pi\sqrt{g^3/\sigma}}}{(g\sigma)^{1/4}}$}
      \rput(1,1){$E_1\approx \frac{\pi^2}{L^2}$}
      \rput(4.5,0.8){$E_1\approx -\frac{\sqrt{g\sigma}}{\pi}\ln\frac{\tilde{L}'}{\ln^2\frac{\pi\sqrt{g^3/\sigma}}{\ln \tilde{L}'}}$}
      \rput(7.8,1){$-g^2$}
      \rput(10.3,1){$E_1\approx-\frac{\sqrt{g\sigma}}{\pi}\ln\tilde{L}\sqrt{\ln\tilde{L}}$}
    \end{pspicture}
    \caption{\it Illustration of regimes for the typical groundstate
      energy $E_1$ with increasing system sizes $L$ in the large
      supersymmetric noise limit $g^3\gg \sigma$.}
    \label{fig:largeggs}
  \end{figure}
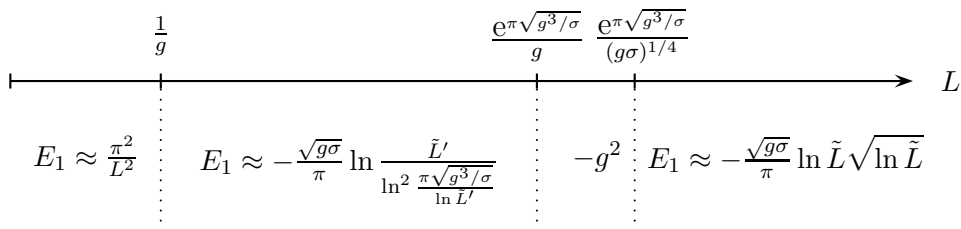

  \begin{figure}[h]
    \centering
    \begin{pspicture}(13,3.4)
      \psline[arrowsize=5pt]{->}(0,2)(12,2)
      \psline(0,1.9)(0,2.1)
      \psline(2,1.9)(2,2.1)\psline[linestyle=dotted](2,1.9)(2,0)
      \psline(6,1.9)(6,2.1)\psline[linestyle=dotted](6,1.9)(6,0)
      \psline(7.3,1.9)(7.3,2.1)\psline[linestyle=dotted](7.3,1.9)(7.3,0)
      \rput(12.5,2){$L$}
      \rput(2,2.6){$\sigma^{-1/3}$}
      \rput(6,2.6){$\frac{\EXP{\frac{8}{3}\sqrt{\sigma/g^3}}}{\sigma^{1/3}}$}
      \rput(7.4,2.6){$\frac{\EXP{\pi\sqrt{\sigma/g^3}}}{(g\sigma)^{1/4}}$}
      \rput(1,1){$E_1\approx \frac{\pi^2}{L^2}$}
      \rput(4.,1){$E_1\approx -\left(\frac{8\sigma}{3}\ln\tilde{L}''\right)^{2/3}$}
      \rput(6.8,1){$-\frac{\sigma}{g}$}
      \rput(9.3,1){$E_1\approx -\frac{\sqrt{g\sigma}}{\pi}\ln \tilde{\mathcal{L}}\sqrt{\ln\tilde{\mathcal{L}}}$}
    \end{pspicture}
    \caption{\it Illustration of regimes for the typical groundstate
      energy $E_1$ with increasing system sizes $L$ in the small
      supersymmetric noise limit $g^3\ll\sigma$. 
      We recall that 
      $\tilde{\mathcal{L}}=\tilde{L}\EXP{-\frac\pi4\sqrt{\sigma/g^3}}$.}
    \label{fig:smallggs}
  \end{figure}
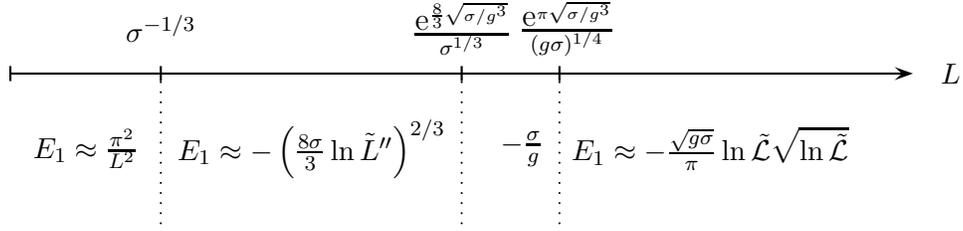

\end{itemize}


\section{Localization}
\label{sec:loc}

Up to now we have concentrated ourselves on the spectral properties of the
random Hamiltonian, however the most stricking property of Hamitonians with
random potentials is the localization of their wavefunctions. In a typical
situation, for example if we consider the Hamiltonian
$H_\mathrm{scalar}=-\frac{\D^2}{\D{}x^2}+V(x)$ where $V(x)$ is random with
short range correlations, one should distinguish two regions in the spectrum~:
in the low energy regime, lowest energy states are those trapped by deep wells
of the potential. The nature of the trapping depends on the statistical
properties of $V(x)$ (Gaussian white noise, low density of repulsive or
attractive impurities, etc). This kind of localization is rather natural. It
is correlative to a rarefaction of states reflected in the Lifshits
exponential tail of the IDoS (section~\ref{sec:Lifshits}). 
On the other hand, in the high energy range ($E\gg$ disorder) the
phenomenon of Anderson localization \cite{And58} takes place~: in a regime
where the static potential is {\it a priori} perturbative, due to interferences
between the extremely large number of scattering paths, the wave functions
decrease exponentially over distances larger than the Fermi wavelength, a
nonperturbative effect. Whereas in the 3d situation a delocalization
(Anderson) transition occurs by tuning the strength of the disordered
potential \cite{AbrAndLicRam79}, the 1d case is particular since all states
are localized \cite{MotTwo61}, a statement rigorously proved in
Refs.~\cite{GolMolPas77,PasFig78}. The problem of 1d Anderson localization has
been reformulated and reexamined in many works (see for example
\cite{Gog82,LifGrePas88,Luc92,Pen94}). As we mentioned in the introduction the
random supersymmetric Hamiltonian presents particular localization properties
since the low energy Dyson singularity of the IDoS \cite{Dys53} is accompanied
by a delocalization transition \cite{BouComGeoLeD90}. These features are strongly
related to the (super)symmetry of the Hamiltonian. In this section we will
examine how the localization picture is modified by breaking  the
supersymmetry in the Hamiltonian (\ref{eqn:hamiltonian}).

Information on localization of wave functions can be obtained by considering
different variables. The most transparent formulation is probably provided by
considering the variables $(\theta,\,\xi)$ of the phase formalism.
Localization length $\ell_\mathrm{loc}$ is related to the damping 
rate of the envelope of the wave function. Therefore we can define the
localization length by analyzing the solution of the Cauchy problem~: from
Eqs.~(\ref{phaseform1},\ref{phaseform2}) we take as a definition the relation
$1/\ell_\mathrm{loc}=\gamma=\lim_{x\to\infty}\frac{\xi(x)}{x}$,
where $\gamma$ is the Lypaunov exponent (note
that we can omit the disorder 
averaging in this definition thanks to self averaging of this process)
\footnote{
  This picture suggests that the wave function behaves roughly as
  $\psi(x)\sim\EXP{\pm\gamma{}x}\times(\mathrm{oscillations})$, however one
  should keep in mind that such a simple picture is dangerous since it forgets
  the important fact that the argument of the exponential, $\xi(x)$, presents
  large fluctuations increasing like $\sqrt{x}$ (fluctuations of $\xi(x)/x$
  vanish for $x\to\infty$, but not those of $\xi(x)$). 
  The envelope of the wave function is an
  exponential of a drifted Brownian motion, what can have important
  consequences~\cite{TexCom99}~; neglecting this important feature can lead to
  wrong conclusion, like in Ref.~\cite{Sor87}.
}.
It is interesting to emphasize that this definition of the localization length
is extracted from the solutions $\psi(x;E)$ of the Cauchy problem, and not
from the real wave functions $\varphi_n(x)$ (solution of the Sturm-Liouville
problem). In other terms the Lyapunov exponent gives a good estimate of
the localization length of $\varphi_n(x)$ if the statistical properties of the
envelope of the solution of the Schr\"odinger equation is not affected when
imposing the second boundary condition. In the high energy limit where
processes $\theta(x)$ and $\xi(x)$ rapidly decorrelate \cite{AntPasSly81} this
is not a problem, however it is not obvious that this holds in any situation
(in particular for the supersymmetric Hamiltonian $H_\mathrm{susy}$, the
Lyapunov exponent does not seem to give a fully satisfactory
information as pointed out in 
the conclusion of Ref.~\cite{Tex00}).

Since the analysis provided in the previous sections is based on the study of
the dynamics of $z(x)$ or $\varphi(x)=\argsinh(\sqrt{{4g}/{\sigma}}\,z(x))$, we
will extract the localization length from the statistical properties of these
stochastic processes. We will derive several formulae and use the most adapted
in the various regimes. Let us recall that the simplest expression of the
Lyapunov exponent is given by the average of the Ricatti
variable~\cite{LifGrePas88}~:
\begin{equation}
  \label{Gamma0}
  \gamma(E) = \mean{ \frac{\psi'(x;E)}{\psi(x;E)} }= \smean{z} + \mean{\phi}
\end{equation}
As in the previous sections we consider here the case $\mean{\phi}=0$. Together with the expression of the stationary distribution $T(z)$, this
immediatly gives the Lyapunov exponent. Note that since $T(z)\simeq{}N(E)/z^2$
for $|z|\to\infty$ (Rice formula), the expression must be understood as
$\gamma=\int_\RR\D{z}\,z\,\frac{T(z)-T(-z)}{2}$ in order to deal with a well
defined integral.
We can also avoid this problem by deriving other formulae, what we do now.

\hspace{0.25cm}

\noindent{\bf Positive part of the spectrum~:}
\mathversion{bold}$E=+k^2$.--\mathversion{normal} We rewrite
the two SDE (\ref{pf1},\ref{pf2}) for phase and envelope as~:
\begin{eqnarray}
  \D\theta &=& 
        k\D x - \frac{\sqrt{\sigma}}{k} \sin^2\theta\, \D W_1(x)
             + \sqrt{g}\,\sin2\theta\, \D W_2(x)   
  \hspace{0.25cm}\mbox{(Stratonovich)} \\
  \D\xi &=& 
            \frac{\sqrt{\sigma}}{2k} \sin2\theta\, \D W_1(x) 
              - \sqrt{g}\,\cos2\theta\, \D W_2(x)  
 \hspace{1.4cm}\mbox{(Stratonovich)}
\end{eqnarray}
where $W_1(x)$ and $W_2(x)$ are two normalized independent Wiener processes.
Since the Lyapunov exponent is related to $\mean{\xi(x)}$ we connect these
Stratonovich-SDE to some Ito-SDE and use the fact that with this latter
prescription, random process and noise are decorrelated at equal ``time''~:
\begin{eqnarray}
  \D\theta &=& 
        \left(k +\frac{\sigma}{2k^2}\sin^2\theta\sin2\theta 
         + \frac{g}{2}\sin4\theta\right)  \D x 
    - \frac{\sqrt{\sigma}}{k} \sin^2\theta\, \D W_1
             + \sqrt{g}\,\sin2\theta\, \D W_2 
  \hspace{0.25cm}\mbox{(Ito)} \\
  \D\xi &=& \left(-\frac{\sigma}{2k^2}\sin^2\theta\cos2\theta 
         + g\sin^22\theta  \right)  \D x +
            \frac{\sqrt{\sigma}}{2k} \sin2\theta\, \D W_1
              - \sqrt{g}\,\cos2\theta\, \D W_2
 \hspace{0.5cm}\mbox{(Ito)}
\end{eqnarray}
We immediatly obtain the following expression~:
\begin{equation}
  \label{GammaTheta}
  \gamma = \frac{\D \mean{\xi}}{\D x}
  = -\frac{\sigma}{2k^2}\smean{\sin^2\theta\cos2\theta }
    +g\smean{ \sin^22\theta }
\end{equation}
where averaging is realized with the stationary distribution.
This relation is similar in spirit to the one derived in Ref.~\cite{HanLuc89}
for $H_\mathrm{scalar}$.
This equation, with the distribution (\ref{disTheta}), gives another explicit
expression for the Lyapunov exponent.
The Lyapunov exponent can also be expressed in term of the distribution
(\ref{disRicatti})
\begin{equation}
    \label{GammaRicatti}
  \gamma = \frac\sigma2
  \mean{
    \frac{E+\left(\frac{8Eg}{\sigma}-1\right)z^2}{(E+z^2)^2}
  }
\end{equation}
or the distribution
(\ref{disVarphi})
\begin{equation}
  \label{Gamma}
  \gamma = 2g
  \mean{\Upsilon_{4Eg/\sigma}(\varphi)}
  \hspace{0.5cm}\mbox{with}\hspace{0.5cm}
  \Upsilon_A(\varphi) \eqdef \frac{A+(2A-1)\sinh^2\varphi}{(A+\sinh^2\varphi)^2}
\end{equation}
Note that the expressions (\ref{GammaTheta},\ref{GammaRicatti},\ref{Gamma})
are valid for $E>0$ and are note appropriate to study the limit $E\to0$~: for
example the equation with the Ricatti variable would take the absurd form
``$\gamma=-\frac\sigma2\smean{\frac1{z^2}}$'' (absurd since $T(z)$ is regular at
$z=0$). The origin of the problem can be understood from (\ref{eqT}) that
shows that in the limit $E\to0$, the two terms
$T(z)=\frac{N(E)}{z^2+E}-\frac{\beta(z)}{2(z^2+E)}\frac{\D}{\D{}z}[\beta(z)T(z)]$
cannot be considered separately. A more detailed discussion is given in appendix \ref{sec:lyapunov}.
\vspace{0.25cm}

\noindent{\bf Band center.--}
In this regime, due to the previous remark, we start from
$\gamma=\smean{z}=\sqrt{\frac{\sigma}{4g}}\smean{\sinh\varphi}$. Using the
fact that the approximate expression of the distribution is symmetric for
$|\varphi|\gtrsim\Phi_0$, we write
$\smean{\sinh\varphi}\simeq\int_{-\Phi_0}^{+\Phi_0}\D\varphi\,\mathcal{P}(\varphi)\,\sinh\varphi$.
We obtain
\begin{equation}
  \boxed{
   \gamma( 0 ) \simeq  \frac{4g}{\ln(16 g^3/\sigma)}
  }
  \hspace{0.25cm}\mbox{for }
  |E|\ll\sqrt{g\sigma}
\end{equation}
Note however that the multiplicative factor $4$ is directly related to our
definition of $\Phi_0$ separating regions where deterministic force and
Langevin force dominates~: $|U'(\Phi_0)|=4g$. Therefore, in this derivation,
the factor $4$ is arbitrary. However the replica method of section
\ref{sec:replica} will predict the same prefactor. We won't have the same
problem for the other regime since we will use the formula (\ref{Gamma})
instead of~(\ref{Gamma0}).

This result shows that even a tiny $\sigma\to0$ scalar noise is sufficient to
lift the delocalization transition of the supersymmetric Hamiltonian.

\vspace{0.25cm}

\noindent{\bf Intermediate energies~:}
\mathversion{bold}$\sqrt{g\sigma}\ll{E}\ll{g}^2$\mathversion{normal}.-- 
The function $\Upsilon_A(\varphi)$ presents two symmetric peaks centered on
$\varphi\simeq\pm\frac12\ln(4A)$. Note that
$\frac12\ln(4A)=\frac12\ln(16Eg/\sigma)=\frac12(\Phi_0+\Phi_E)$. We remark
that, for $A=\frac{4Eg}{\sigma}\gg1$, we have
$\int_0^\infty\D\varphi\,\Upsilon_A(\varphi)\simeq1$ and
$\int_0^\infty\D\varphi\,\varphi\,\Upsilon_A(\varphi)\simeq\frac12\ln(4A)$
(these equalities are already excellent for $A=0.5$). Therefore, using the
approximate form of the distribution derived above, we can write
\begin{equation}
  \gamma\simeq\frac{2g}{(\Phi_0-\Phi_E)^2}
  \left[
    \int_{-\infty}^0\D\varphi\,(\varphi+\Phi_0)\Upsilon_A(\varphi)
    +
    \int_0^\infty\D\varphi\,(\varphi-\Phi_E)\Upsilon_A(\varphi)
  \right]
  =\frac{2g}{\Phi_0-\Phi_E}
\end{equation}
Therefore we recover the result obtained for the supersymmetric Hamiltonian alone
\cite{BouComGeoLeD90}
\begin{equation}
  \boxed{
   \gamma( E ) \simeq  \frac{2g}{ \ln(16g^2/E) }
  }
  \hspace{0.25cm}\mbox{for }
  \sqrt{g\sigma}\ll{E}\ll{g}^2 
\end{equation}

\vspace{0.25cm}

\noindent{\bf High energy limit.--}
In the high energy limit the distribution of the phase $\theta$ is almost
flat, therefore using (\ref{GammaTheta})
\begin{equation}
  \gamma(E\to+\infty) \simeq \frac{\sigma}{8E} + \frac{g}{2} 
  = \gamma_\mathrm{scalar} + \gamma_\mathrm{susy}
\end{equation}
where $\gamma_\mathrm{scalar}\simeq\frac{\sigma}{8E}$ and
$\gamma_\mathrm{susy}\simeq\frac{g}{2}$ are the high energy Lyapunov
exponents for $H_\mathrm{scalar}=-\frac{\D^2}{\D{}x^2}+V(x)$ and
$H_\mathrm{susy}$, respectively.

For $E\to\infty$ the localization length saturates to
$\ell_\mathrm{loc}\simeq2/g$. The high energy wave functions present
rapid oscillations over a scale $1/k$ exponentially damped on a larger
scale~$2/g$.

\vspace{0.25cm}

\noindent{\bf Negative part of the spectrum~:}
\mathversion{bold}$E=-k^2$.--\mathversion{normal} As we have seen above,
compare to the SDE for $E=+k^2$, the SDE for the variable $\xi$ for $E=-k^2$
receives an additional term $k\sin2\theta$, therefore
\begin{equation}
  \gamma 
  = k\smean{\sin2\theta} 
   -\frac{\sigma}{2k^2}\smean{\sin^2\theta\cos2\theta }
    +g\smean{ \sin^22\theta }
\end{equation}
In the limit $E\to-\infty$ the phase is trapped at $\theta\simeq\pi/4$ (this
is related to trapping of $\varphi$ by the local minimum of potential
$U(\varphi)$ at $\varphi_+$), therefore
\begin{equation}
  \gamma(E\to-\infty) \simeq \sqrt{-E} + g 
\end{equation}
This increase of the Lyapunov exponent reflects that the low energy
wave functions are sharply peaked around deep wells of the potential.

\section{Replica method}
\label{sec:replica}

In this section we derive analytic expressions for the IDoS and the Lyapunov
exponent by using the replica method. The computation consists of a slight
variant of the method used in Ref.~\cite{BouComGeoLeD90}, which leads to
hypergeometric functions, generalizing the Bessel and Airy functions appearing
in the pure supersymmetric and pure scalar potential problem
respectively~\cite{LeD07}. Therefore, we only sketch the main lines and refer
to the Ref.~\cite{BouComGeoLeD90} for details.

We consider the Hamiltonian (\ref{eqn:hamiltonian}) with $V(x)$ and $\phi(x)$
two uncorrelated Gaussian white noises, in the more general case where
$\smean{\phi(x)}$ is finite~: $V(x)=\sqrt{\sigma}\eta(x)$ and
$\phi(x)=\mu\,g+\sqrt{g}\tilde{\eta}(x)$ ($\eta(x)$ and $\tilde{\eta}(x)$ with
$\mu>0$ are two uncorrelated normalized Gaussian white noises of zero means).
As mentioned above, the problem of $\delta$-correlations between the noises
may be mapped on the uncorrelated case (see appendix \ref{app:corrnoises}).
The spectral properties of $H$ are encoded in the Green's function $G(x,y;E)$
given by the matrix element
\begin{equation}
  G(x,y;E) = \big(x\big| \frac1{E-H} \big|y\big)
  =\sum_{\alpha}\frac{\Psi_\alpha(x)\Psi_\alpha^\ast(y)}{E-E_\alpha}
\end{equation}
in position space. Here $\Psi_\alpha$ denotes the eigenfunction associated to
the energy level $E_\alpha$, and the sum runs over all states $\alpha$.
According to Thouless' formula, average with respect to disorder
$\smean{G(x,x;E)}$ of the Green's function at equal points yields the
derivative of Lyapunov exponent as a function of $E$ \cite{Tho72}. Analytic
continuation $E\to E-\I0^+$ allows to write
\begin{equation}
    \label{eqn:greenep}
    \langle G(x,x;E-\I0^+) \rangle = \gamma'(E) + \I \pi\rho(E)
\end{equation}
where $\rho(E)$ is the density of states per unit length.

\mathversion{bold}
\subsection{The $n$-replica Hamiltonian}
\mathversion{normal}

We shall make use of the replica trick in order to compute the averaged
equal-point Green's function \eqref{eqn:greenep} (see for example
\cite{ItzDro89}). To this end, we introduce an auxiliary $n$-component field
$\chi=(\chi^1,\dots,\chi^n)$ and rewrite $\langle G(x,x;E)\rangle$ in terms of
a Gaussian path integral with respect to~$\chi$:
\begin{align}
\langle (x|(H-E)^{-1}|x)\rangle 
  &= \frac{1}{L}\int_{-L/2}^{+L/2}\D x\,\langle (x|(H-E)^{-1}|x)\rangle\nonumber\\
  &= \frac{1}{L}\lim_{n\to 0}\frac\partial{\partial n}
      \int_{-L/2}^{+L/2}\D x\,\int\mathcal{D}\chi\:\chi(x)^2
      \left\langle
        \exp\left(-\frac{1}{2}\int_{-L/2}^{+L/2}\D
          y\,\chi(y)(H-E)\chi(y)\right)
      \right\rangle\nonumber\\
  &= \frac{2}{L}\frac{\partial}{\partial E}\lim_{n\to 0}\frac\partial{\partial n}
   \int\mathcal{D}\chi
   \left\langle
     \exp\left(-\frac{1}{2}\int_{-L/2}^{+L/2}\D y\,\chi(y)(H-E)\chi(y)\right)
   \right\rangle
   \label{eqn:pathint}
\end{align}
Notice that the first line makes explicitly use of translation invariance after average with
respect to disorder. The limit $n\to 0$ eliminates the residual determinant
from path integration with respect to $\chi$. We thus are interested in
the $n$-replica partition function
\begin{equation}
  {\cal Z}_n 
  = \int\mathcal{D}\chi
     \left\langle
       \exp\left(-\frac{1}{2}\int_{-L/2}^{+L/2}\D x
         \,\chi(x)(H-E)\chi(x)\right)
     \right\rangle
  =\int\mathcal{D}\chi
  \exp\left(-\int_{-L/2}^{+L/2}\D x\,{\cal L}(\chi,\dot{\chi})\right)
  \label{eqn:partition}
\end{equation}
where the average over disorder has lead to the Lagrangian
\begin{align}
  {\cal L}(\chi,\dot{\chi}) &= 
  \frac{1}{2}\dot{\chi}^2 
  -\frac{g\,\left(\chi\cdot\dot{\chi}\right)^2}{2(1+ g\chi^2)}
  +\frac{\mu^2g^2\chi^2}{1+g\chi^2}-\frac{E}2\,\chi^2-\frac\sigma8\,(\chi^2)^2
  +\frac{1}{2}\delta^{(n)}(0)\,\ln\det(1+g\chi^2).
\end{align}
As the formula suggests, we abbreviate $\chi^2=\sum_i(\chi^i)^2$ and the
scalar product $\chi\cdot\eta=\sum_i\chi^i\eta^i$. Rewriting $\mathcal L$
as
\begin{equation}
  \mathcal{L}=\frac{1}{2}\sum_{i,j=1}^n
  \eta_{ij}(\chi)\dot{\chi}^i\dot{\chi}^j+V(\chi),
  \qquad
  \eta_{ij}(\chi) = \delta_{ij}-\frac{g\,\chi_i\chi_j}{1+g\chi^2}.
\label{eqn:lagrangian}
\end{equation}
shows that the Lagrangian describes the motion of a point particle in an
$n$-dimensional curved space with metric $\eta_{ij}$. The potential is given
by
\begin{equation}
  V(\chi)=\frac{\mu^2g^2\chi^2}{1+g\chi^2}-\frac{E}2\,\chi^2
  -\frac\sigma8\,(\chi^2)^2  +\frac{1}{2}\delta^{(n)}(0)\,\ln\det(1+g\chi^2)
  \label{eqn:lpot}
\end{equation}
The contact term $\delta^{(n)}(0)$ may be eliminated by introducting an
auxiliary field $\Sigma = \sqrt{1+g\chi^2}$ and rewriting the functional
integration measure as
$\mathcal{D}\chi\mathcal{D}\Sigma\,\delta(\Sigma^2-g\chi^2-1)$~; following
\cite{BouComGeoLeD90} this term will not be considered in the sequel. We
recognize an $\sigma$-model with symmetry group $O(n,1)$. In one spatial
dimension we may transform it to a quantum-mechanical problem in
$n$-dimensions where $x$ plays the role of time. Hence we must identify a
proper Hamiltonian $\mathcal{H}$ related to $\mathcal{L}$ and study its
spectrum. $\mathcal H$ acts on a Hilbert space with inner product
\begin{equation}
  (\Phi,\Psi) = 
  \int_{\mathbb{R}^n}\D^n\chi\,\sqrt{\det{\eta}}\:
  \Phi^\ast(\chi)\Psi(\chi) 
  = \int_{\mathbb{R}^n}\frac{\D^n \chi}{\sqrt{1+g\chi^2}}\: 
  \Phi^\ast(\chi)\Psi(\chi)
\end{equation}
and its eigenvalues $\mathcal{E}_\nu(n)$ and eigenfunctions
$\Psi_\nu(\chi)$ are given by the solutions of
$\mathcal{H}\Psi_\nu(\chi) = \mathcal{E}_\nu(n)\Psi_\nu(\chi)$ with
$||\Psi_\nu||^2=(\Psi_\nu,\Psi_\nu)< \infty$. Since the derivation of
$\mathcal{H}$ is very much like in \cite{BouComGeoLeD90} we only state the
result. From \eqref{eqn:lagrangian} and \eqref{eqn:lpot} we find the
$n$-replica Hamiltonian
\begin{equation}
  \mathcal{H} = 
  \frac{1}{2}
  \left(
    -\Delta +
    (1-n)\chi\cdot\nabla -(\chi\cdot\nabla)^2 +\frac{n}{2}
   -\frac{1}{4}\frac{\chi^2}{1+\chi^2}
  \right)+V(\chi).
\end{equation}
The angular eigenstates are given by the Gegenbauer polynomials
$C_\ell^{n/2-1}(\cos\theta)$ where $\ell$ denotes the main angular quantum
number. After separation of the angular part, we are left with the radial part
of the Hamiltonian that depends only on the modulus $\rho=\sqrt{\chi^2}$:
\begin{align}
  \label{eqn:radial}
  \mathcal{H}_{\mbox{r}} =&  
  -\frac{1}{2}(1+g\rho^2)\frac{\partial^2}{\partial \rho^2}
  -\frac{n-1}{2\rho}\frac{\partial}{\partial \rho}
  -\frac{ng\rho}{2}\frac{\partial}{\partial \rho}
 +\frac{\ell(\ell+n-2)}{2\rho^2}+\frac{gn}{4}
  +\frac{\mu^2g^2\rho^2}{1+g\rho^2}-\frac{E}2\,\rho^2-\frac\sigma8\,\rho^4
\end{align}

\subsection{The groundstate~: Lyapunov exponent and IDoS}

In the limit $L\to \infty$ we expect that the path integral
\eqref{eqn:partition} has a leading term $\exp[-L\mathcal{E}_\mathrm{G}(n)/2]$
where $\mathcal{E}_\mathrm{G}(n)/2$ corresponds to the ground-state energy of
the Hamiltonian $\mathcal{H}$. Combining \eqref{eqn:greenep},
\eqref{eqn:pathint} and \eqref{eqn:partition} we conclude that
\begin{equation}
  \gamma(E)+\I\pi N(E)+ \text{const.}=-\frac{2}{L}\left.
   \frac{\partial {\mathcal Z}_n}{\partial n}\right|_{n=0} 
  =  \left. \frac{\partial \mathcal{E}_\mathrm{G}(n)} {\partial n}\right|_{n=0}.
\end{equation}
As above, $N(E)$ denotes the integrated density of states per unit length, and
$\gamma(E)$ the Lyapunov exponent. The constant must be chosen in order to
ensure correct asymptotic behaviour $E\to\pm\infty$ (in particular
$N(E\to-\infty)=0$). We shall discuss this problem below. Following the spirit
of the replica method, we analytically continue
$\mathcal{E}_\mathrm{G}(n)=n\mathcal{E}_0+n^2\mathcal{E}_1+\cdots$ and thus
identify
\begin{equation}
  \gamma(E) + \I \pi N(E) = \mathcal{E}_0 + \text{const.}
\end{equation}
We now compute $\mathcal{E}_0$ for the Hamiltonian (\ref{eqn:radial}). We expect the
ground state to be an s-wave state with total angular momentum $\ell=0$. Changing
variables to $\xi^2 = 1+g\rho^2$ in (\ref{eqn:radial}) leads to the Hamiltonian
\begin{equation}
\mathcal{H}_{\mbox{r}} = \frac{g(\xi^2-1)}{2}\,\mathcal{T}
 +\frac{ng}{2}\left(\frac{1}{2}-\xi\frac{\partial}{\partial\xi}\right),
 \quad \mbox{with } 
  \mathcal{T} = -\frac{\partial^2}{\partial \xi^2} 
  -\frac{E}{g^2}-\frac{\sigma}{4g^3}(\xi^2-1)+\frac{\mu^2-1/4}{\xi^2}.
\label{eqn:diffeqns}
\end{equation}
Consequently we must solve the equation
$\mathcal{H}_{\mbox{r}}\Psi=\frac12\mathcal{E}_\mathrm{G}(n)\Psi$ for the the
ground state wave function. As for the eigenvalue $\mathcal{E}_\mathrm{G}(n)$,
we expand the ground-state wave function into a power series with respect to
$n$: $\Psi=\Psi_0+n\Psi_1+\cdots$ This yields an infinite system of coupled
differential equations whose first two members are
\begin{align}
\label{eqn:diffeqns2}
  \mathcal{T}\Psi_0=0\qquad
  \text{and}\qquad\frac{g(\xi^2-1)}{2}\mathcal{T}\Psi_1(\xi)
  +\frac{g}{2}\left(\frac{1}{2}-\xi\frac{\partial}{\partial\xi}\right)\Psi_0(\xi)
  =\frac{\mathcal{E}_0}{2}\Psi_0(\xi).
\end{align}
Since we seek for a normalizable groundstate wavefunction in the limit $n\to
0$ we have to find a square-integrable solution of $\mathcal{T}\Psi_0(\xi)=0$.
Applying the limit $\xi\to1$ in \eqref{eqn:diffeqns2}, we finally may relate
$\Psi_0$ to the eigenvalue
\begin{equation}
  \mathcal{E}_0 = g
  \left.
     \left(\frac{1}{2}-\frac{\xi}{\Psi_0(\xi)}\frac{\partial\Psi_0(\xi)}{\partial\xi}\right)
  \right|_{\xi=1}.
\end{equation}
The solution is given in appendix \ref{sec:sol} and yields the wave function $\Psi_0$
\begin{equation}
  \Psi_0(\xi) = \exp\left(-\frac{\I\xi^2}{4}\sqrt{\frac{\sigma}{g^3}}\right)\:
  \xi^{\mu+1/2}\: 
  U\left(
      \frac{\mu+1}{2} 
      +\frac{\I}{2}
        \left(\frac{E}{\sqrt{\sigma g}}-\frac{1}{4}\sqrt{\frac{\sigma}{g^3}}\right),
     \mu+1,
     \frac{\I\xi^2}{2}\sqrt{\frac{\sigma}{g^3}}
   \right)
\end{equation}
where $U(a,b,z)$ denotes the second confluent hypergeometric function
\cite{AbrSte64}. Therefore, $\mathcal{E}_0$ takes the value
\begin{equation}
  \label{eqn:e0}
  \boxed{
  \mathcal{E}_0 = -\mu\, g
  -\frac{\I}{2}\sqrt{\frac{\sigma}{g}}
   \left(
      1-
     \frac{2a\,U(a+1,\mu+2,\I\sqrt{\sigma/4g^3})}
          {(\mu+1)\;U(a,\mu+1,\I\sqrt{\sigma/4g^3})}
  \right)
  }
\end{equation}
where we have introduced
\begin{equation}
  \label{eqn:ab}
  a= \frac{\mu+1}{2} + \frac{\I}{2}\left(\frac{E}{\sqrt{\sigma g}}
    -\frac{1}{4}\sqrt{\frac{\sigma}{g^3}}\right)
  \:.
\end{equation}
The imaginary part can be extracted by using the Wronskian
\eqref{eqn:wronskian} of $\Psi_0(\xi)$ and its complex conjugate~:
\begin{equation}
  \label{newresult}
  \boxed{
  N(E) = \frac{g}\pi
  \left(\frac{4g^3}{\sigma}\right)^{\frac\mu2}\,
  \frac{\exp (\pi \im a)}{|U(a,\mu+1,\I\sqrt{\sigma/4g^3})|^2}
  }.
\end{equation}
We have obtained a compact expression, that can be used more
conveniently than the double integral (\ref{IDoS}) in order to plot the IDoS.

\begin{figure}[htpb]
  \centering
  \begin{tabular}{cc}
  \includegraphics[width=0.475\textwidth]{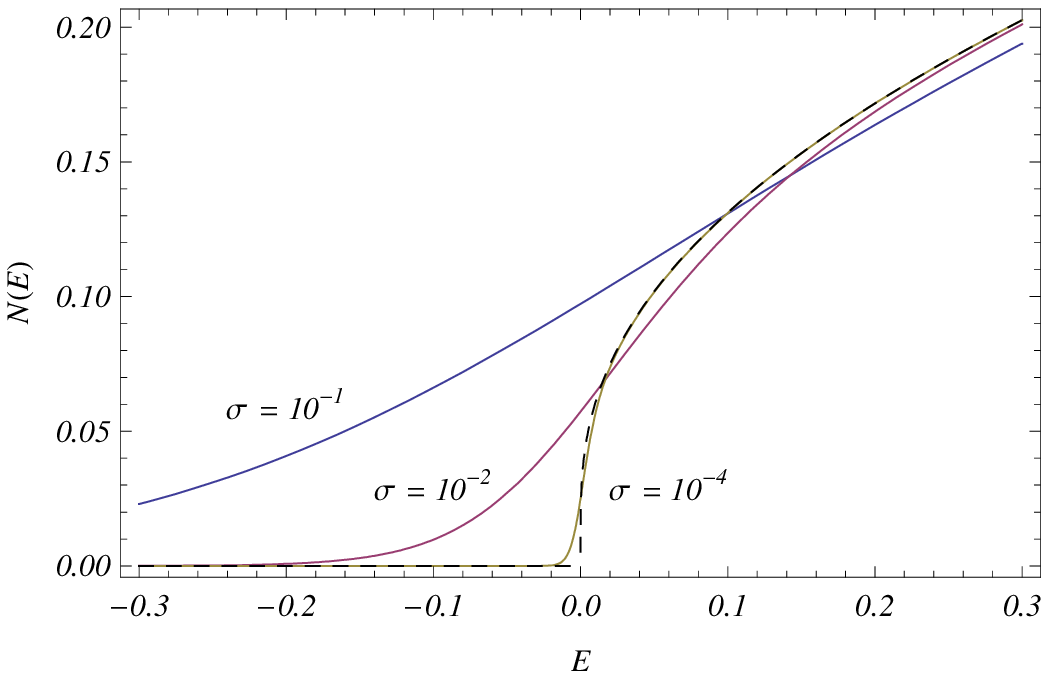}&
  \includegraphics[width=0.475\textwidth]{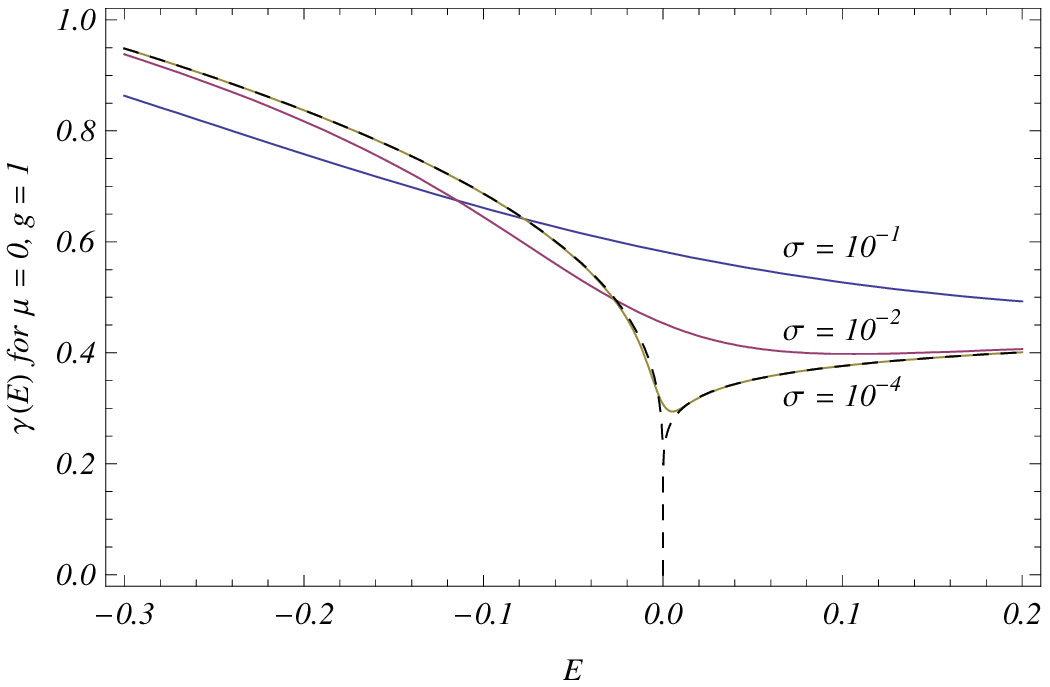}\\
  (a) & (b)
  \end{tabular}
  \caption{\it IDoS (left) and Lyapunov exponent (right) for $g=1$ and
    $\mu=0$ for various values of $\sigma$.
    Delocalization transition at $E=0$ for $\sigma=0$ (dashed lines) is suppressed even by a
    tiny scalar potential.}
  \label{fig:deloc}
\end{figure}

Eqs.~(\ref{eqn:e0},\ref{eqn:ab},\ref{newresult}) provide an exact solutions
for the Lyapunov exponent $\gamma(E)$ as well as the IDoS $N(E)$ for this
model, up to a constant which depends upon $\sigma$, $g$ and $\mu$, and may be
fixed by imposing correct asymptotic behaviours, like
$\lim_{E\to-\infty}N(E)=0$. These results interpolate between the known cases
of white noise potential and the random supersymmetric Hamiltonian.

Let us give an example on how to use (\ref{eqn:e0},\ref{eqn:ab}) to study the
behaviour at $E=0$ for the Sinai case $\mu=0$. We have
\begin{equation}
  a=\frac{1}{2}-\frac{\I}{8}\sqrt{\frac{\sigma}{g^3}}
\end{equation}
Recall that the confluent hypergeometric function $U(a,b,z)$ behaves like
\begin{equation}
  U(a,b,z) \underset{z\to 0}{\sim}\left\{
    \begin{array}{cc}
      \displaystyle{\Gamma(b-1)/\Gamma(a)}\,z^{1-b}, & b > 1\\
      \\
      (\ln z + \psi(a))/\Gamma(a),&b=1
    \end{array}
  \right. .
  \label{eqn:ulimit}
\end{equation}
For small Gaussian noise $\sigma\to 0^+$ we tacitely neglect the small imaginary
part of $a$, leading to further corrections, and find
\begin{align}
  \mathcal{E}_0&
  \underset{E=0}{\approx} + \frac{\I}{2}\sqrt{\frac{\sigma}{g}}
  \left(1-\frac{2}{\displaystyle\ln\left(\frac{\I}{2}\sqrt{\frac{\sigma}{g^3}}\right)
  +\psi(1/2)}\left(\frac{\I}{2}\sqrt{\frac{\sigma}{g^3}}\right)^{-1}\right)
  \\
  &\approx  -\frac{2g}{\displaystyle\left(\ln\left(\frac{1}{2}\sqrt{\frac{\sigma}{g^3}}\right)+\psi(1/2)\right)^2+\frac{\pi^2}{4}}\left(\displaystyle\ln\left(\frac{1}{2}\sqrt{\frac{\sigma}{g^3}}\right)+\psi(1/2)-\frac{\I\pi}{2}\right).
\end{align}
Therefore we obtain the approximate IDoS~:
\begin{equation}
  \boxed{
  N(0) \simeq
  \frac{g}
       {
   \left[\ln\sqrt{g^3/\sigma}+\ln2-\psi(1/2)\right]^2 + \pi^2/4
       } 
  }.
\end{equation}
We have recovered by the replica method the behaviour obtained in the
sections~\ref{sec:spec} and \ref{sec:loc}
\begin{equation}
  \label{eq108}
  N(E=0)\underset{\sigma\to 0}{\sim}
  \frac{g}{\ln^2(g^3/\sigma)}
  \qquad\text{and}\qquad 
  \gamma(E=0)\underset{\sigma\to 0}{\sim} 
  \frac{g}{\ln(g^3/\sigma)} .
\end{equation}
Note however that the next leading order are different (this is not surprising
since the approximation scheme of section~\ref{sec:spec} is quite different).
Nevertheless, (\ref{eqn:e0},\ref{eqn:ab},\ref{newresult}) are less manageable for the intermediate regimes singled out in the previous sections.

Figure \ref{fig:deloc} illustrates $N(E)$ and $\gamma(E)$
for the Sinai's case ($\mu=0$, $g=1$). Any Gaussian noise with $\sigma>0$ lifts
the singular behaviour $N_{\mathrm{susy}}(E)\sim 1/(\ln E)^2$ and
$\gamma_{\mathrm{susy}}(E)\sim1/|\ln E|$ to analyticity in the vicinity of
$E=0$. In particular, as shown on figure \ref{fig:deloc}, any small $\sigma$
shifts the singularity of $\gamma(E)$ to some minimum at some
$E_{\mathrm{min}}>0$.

\vspace{0.25cm}

\begin{figure}[htpb]
  \centering
  \begin{tabular}{cc}
  \includegraphics[width=0.475\textwidth]{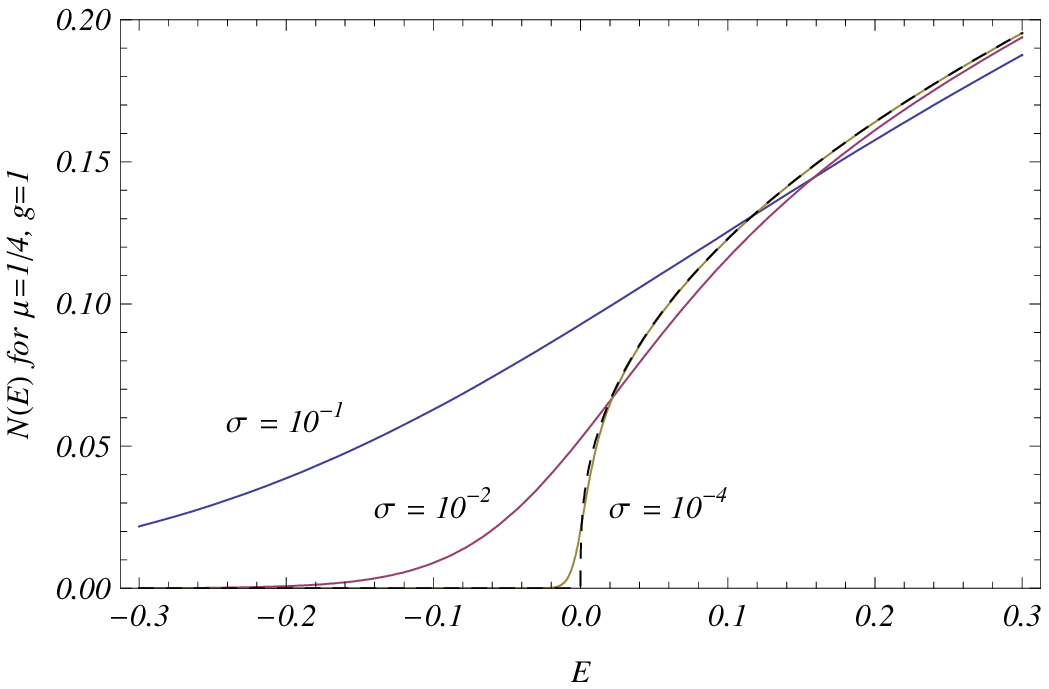}&
  \includegraphics[width=0.475\textwidth]{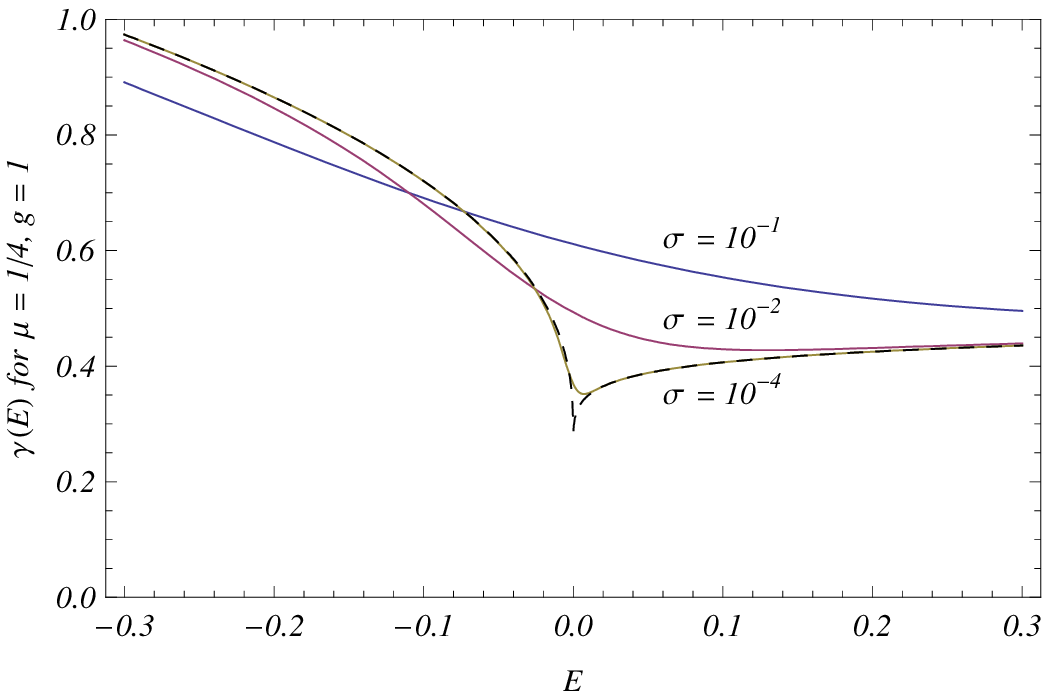}\\
  \includegraphics[width=0.475\textwidth]{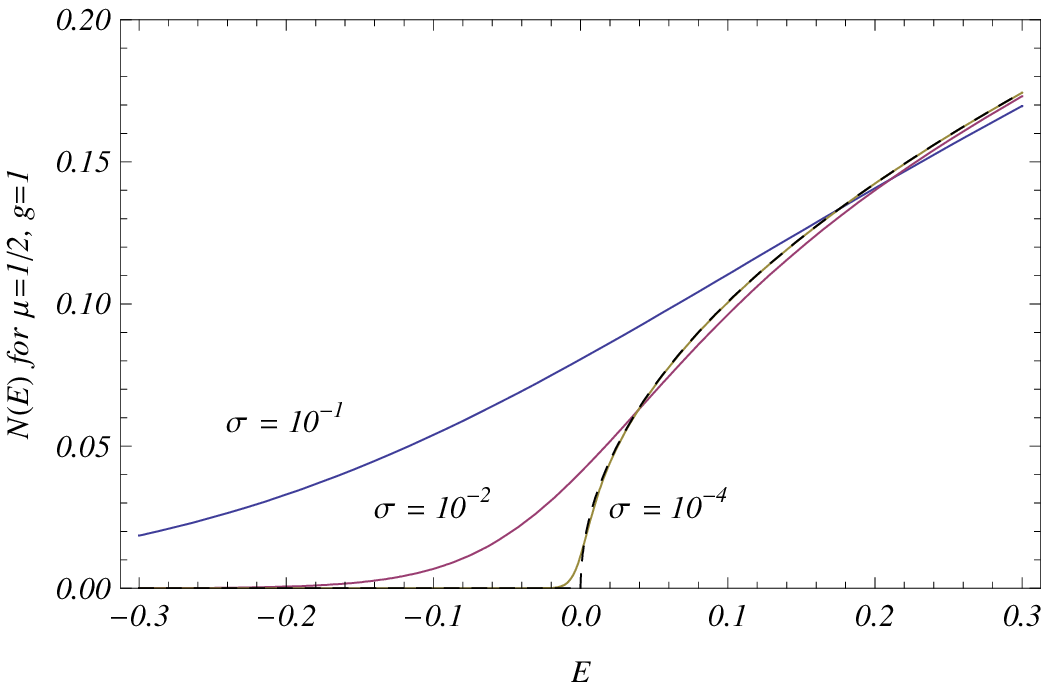}&
  \includegraphics[width=0.475\textwidth]{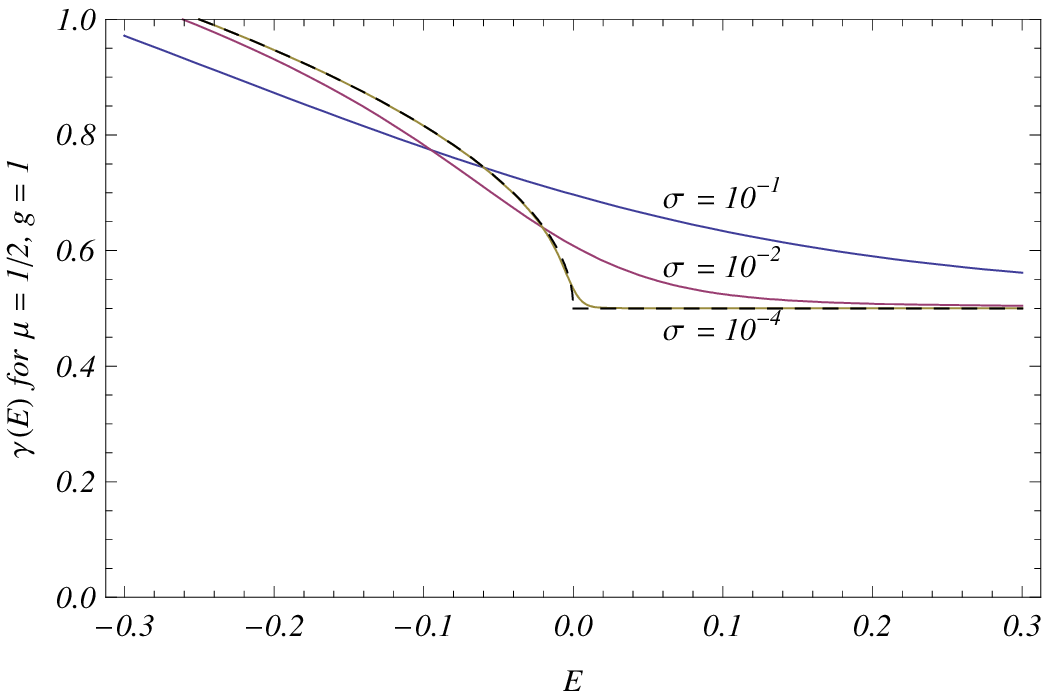}
  \end{tabular}
  \caption{\it $N(E)$ and $\gamma(E)$ for $\mu=1/4$ (top) and $\mu=1/2$
    (bottom). The dashed lines correspond to the pure supersymmetric results for
    $\sigma=0$.}
  \label{fig:muneqzero}
\end{figure}

\mathversion{bold}
\noindent{\bf The case $\mu\neq0$.--}
\mathversion{normal}
It is also interesting to consider the case of finite $\mean{\phi}=\mu\,g$. 
In the absence of the scalar noise $V(x)$ ($\sigma=0$) the power law Dyson
singularity of the IDoS is transformed into a power law behaviour
$N(E)\sim{E}^\mu$. If a tiny scalar noise is introduced a fraction of states
migrates to $\RR^-$~:
\begin{equation}
  \label{eq109}
  \boxed{
  N(E=0) \underset{\sigma\to0}{\simeq} \frac{g}\pi
  \left[
    \frac{\Gamma\left(\frac{\mu+1}{2}\right)}{\Gamma(\mu)}\,
  \right]^2
  \,\left(\frac{\sigma}{4g^3}\right)^{\frac\mu2}
  }
\end{equation}
what we find by straightforward application of \eqref{eqn:ulimit} to
\eqref{newresult}.  
Moreover the feature of smoothing singular behaviour extends to $0<\mu<1/2$.
For $\sigma=0$ we have the non-analytic behaviour
$\gamma_{\mathrm{susy}}(E)\sim \mu\,g+C_{\pm}\,|E|^{\mu}$ with some constants
$C_\pm$ for $E>0$ and $E<0$ respectively. Again, introduction of $\sigma$
shifts this power-law singularity to some minimum of $\gamma(E)$ at small
positive $E_{\mathrm{min}}$, as illustrated on figure \ref{fig:muneqzero}. In
either case, the evaluation of $E_{\mathrm{min}}$ seems to be difficult.
However, it would be interesting to find a physical argument for this
mechanism.


\section{Conclusion}
\label{sec:conclusion}

In this article we have studied spectral and localization properties of a
one-dimensional random Hamiltonian $H=-\frac{\D^2}{\D
  x^2}+\phi(x)^2+\phi'(x)+V(x) = H_{\mathrm{susy}}+V(x)$ which interpolates
between the well-studied examples of random supersymmetric models
$H_{\mathrm{susy}}$ and Halperin's model $H_{\mathrm{scalar}}$. Our analysis
has pointed out a natural competition between the fluctuations of $\phi(x)$ and
$V(x)$. We have identified the important scales that control this
competition for $g^3\gg\sigma$ or $g^3\ll\sigma$, which are the two
largest scales among $\sigma/g$, $\sigma^{2/3}$, $\sqrt{g\sigma}$ and $g^2$.
We have observed that even a small additional scalar noise $V(x)$ lifts the
singular spectral and localization properties of $H_{\mathrm{susy}}$~: the
Dyson singularity of the IDoS and the vanishing of the Lyapunov exponent at
$E=0$ are replaced by smooth behaviours~: a small additional scalar white
noise ($\sigma\to0$) leads to a migration of a fraction
$N(0)\sim{}g/\ln^2(g^3/\sigma)$ of eigenstates to negative values. It is worth
noticing that, $\forall\,\mu$, the zero energy IDoS (\ref{eq108},\ref{eq109})
for $g^3\gg\sigma$ can be obtained by the substitution $E\to\sqrt{g\sigma}$ in
the known expressions for~$\sigma=0$~:
\begin{equation}
  N^{(\sigma\neq0)}(E=0) \sim N^{(\sigma=0)}(E\sim\sqrt{g\sigma}).
\end{equation}
This is a simple consequence of the correct identification of the
crossover energy scales.

Simultaneously to the smoothing of the Dyson singularity, the
delocalization transition of $H_{\mathrm{susy}}$ at $E=0$ disappears
and the Lyapunov exponent takes a finite value
$\gamma(0)\sim\gamma^{(\sigma=0)}(E\sim\sqrt{g\sigma})\sim{}g/\ln(g^3/\sigma)$.
This logarithmic behaviour shows that, in practice (see
figure~\ref{fig:deloc}), even a tiny Gaussian noise $\sigma$ kills the
singularity of the Lyapunov that becomes almost flat
$\gamma(E)\sim{}g$ for all energies for which density of states is
significant.

IDoS and Lyapunov exponent have also been studied in the other
regimes. In particular, how the fraction
$N(0)\sim{}g/\ln^2(g^3/\sigma)$ of states are distributed among
negative energies has been further analyzed~; the precise (Lifshits)
exponential tail of the IDoS has been derived in the various regimes.
It is worth emphasizing that in the lowest part of the spectrum, the
tail involves a competition between the supersymmetric and the scalar
noise, $N(E\to-\infty)\sim\exp(-\frac{\pi}{\sqrt{g\sigma}}|E|)$,
whatever is the largest scale among $g^3$ (supersymmetric noise) and
$\sigma$ (scalar noise).

The study of spectral properties has been completed
by considering the individual distributions of eigenenergies (extreme value
problem). We have shown that these distributions coincide with Gumbel
laws, a consequence of the absence of spectral correlations due to the
strong localization of the wave functions~\cite{Mol81}, like for the scalar
potential alone~; this can be opposed to the
purely supersymmetric case ($\sigma=0$) for which distributions of
eigenenergies are strongly modified in the neighbourhood of the
delocalization transition~\cite{Tex00}.

The study of individual distributions of eigenenergies, that includes
properly finite size (Dirichlet boundary) effects, had allowed
us to identify the critical value $\sigma_c$ of the scalar noise $\sigma$ below
which, for fixed $g$ and $L\gg1/g$, the scalar noise can be
ignored. We have obtained $\sigma_c\sim{}g^3\EXP{-\sqrt{gL}}$.
It is worth noticing that the corresponding value of the $E=0$
Lyapunov exponent (roughly its minimum value) then reads 
$\gamma(0)\sim{}g/\ln(g^3/\sigma_c)\sim\sqrt{g/L}$. This corresponds
to a maximum localization length~$\ell_\mathrm{loc}\sim\sqrt{L/g}\ll{}L$.

It is not too surprising that the additional white noise modifies spectral and
localization properties in the vicinity of the band center (around $E=0$).
However, it is somewhat unexpected that, at any value of $g$ (even in the
limit $g\to0^+$), the noise $\phi(x)$ from the supersymmetric part controls
the spectral properties for $E\to-\infty$, what we have seen on the tail
$N(E)\sim\exp(-\frac{\pi}{\sqrt{g\sigma}}|E|)$ and the distributions of the
lowest energy levels. This feature seems counter-intuitive since the pure SUSY
spectrum is strictly positive so that we would have expected the potential
$V(x)$ to yield the behaviour $N(E)\sim\exp(-\frac{8}{3\sigma}|E|^{3/2})$. We
attribute this behaviour to the singular nature of the supersymmetric
potential $\phi(x)^2+\phi'(x)$ which is also responsible for the saturation of
the Lyapunov exponent at high energies $\gamma(E)\simeq{}g/2$ for
$E\to+\infty$. Part of this picture will change if supersymmetric noise is
replaced by a more regular process with regular correlation function of finite
width and height (see Ref.~\cite{ComDesMon95}).

\vspace{0.25cm}

\noindent{\bf Diffusion in random force field with random annihilation/creation
rates.--}
Finally it is interesting to come back to the analysis of the results in the
context of classical diffusion in random force field with random
annihilation/creation rates.
In order to distinguish more clearly the roles of the force field $\phi(x)$
and the annihilation/creation rates $V(x)$, we consider several situations and
analyze the density of particles $\smean{n(x,t|x,0)}$ at $x$ at time
$t$, when a particle has been released at $x$ initially. Averaging is taken
over the random force field and the random annihilation/creation rates.

\begin{itemize}
\item\mathversion{bold}{\bf For $g=0$ and $\sigma=0$~:}\mathversion{normal}
  It is useful to recall the obvious fact that in the absence of random force
  field and absorption we have $n(x,t|x,0)=\frac1{\sqrt{4\pi\,t}}$.

\item\mathversion{bold}{\bf For $g\neq0$ and $\sigma=0$~:}\mathversion{normal}
  Classical diffusion in a random force field (Sinai problem). Thanks to
  (\ref{Laplace}), the spectral Dyson singularity $N(E)\sim1/\ln^2E$ can be
  connected to large time behaviour~\cite{BouComGeoLeD90}
  \begin{equation}
    \label{anomalousdiff}
    \smean{n(x,t|x,0)} \underset{t\to\infty}{\sim} \frac1{\ln^2t}
  \end{equation}
  much slower than the $1/\sqrt{t}$. This behaviour is related to the
  behaviour $x(t)\sim\ln^2t$ of the typical distance covered by the random
  walker~\cite{BouComGeoLeD90} (see also Ref.~\cite{LeDMonFis99} where many
  interesting properties of the Sinai problem were studied thanks to the
  powerful real space renormalization group method of Ma \& Dasgupta).

\item\mathversion{bold}{\bf For $g=0$ and $\sigma\neq0$~:}\mathversion{normal}
  In order to examine the effect of the annihilation/creation rates that were
  chosen to be zero on average, we first switch off the random force field. Of
  course the number of particles is not conserved for $\sigma\neq0$. In this
  case the spectral Lifshits singularity of the DoS is
  $\rho(E)\simeq\frac{1}{2\pi\sqrt{|E|}}\exp-\frac{8|E|^{3/2}}{3\sigma}$. The
    Laplace transform (\ref{Laplace}) is dominated by negative energy
    contributions. A steepest descent estimation shows that the averaged
    number of returning particles diverges with time as~:
  \begin{equation}
    \smean{n(x,t|x,0)} \underset{t\to\infty}{\simeq}
    \frac1{\sqrt{\pi t}}\, \EXP{+\frac{\sigma^2}{48}t^3}.
  \end{equation}
  We emphasize that this increase of the averaged density cannot be
  compensated by a finite mean value of the annihilation rates $\smean{V}>0$ that
  would only add a $\EXP{-\smean{V}t}$ to this result.

\item\mathversion{bold}{\bf For $g\neq0$ and $\sigma\neq0$~:}\mathversion{normal}
  Finally we consider the case of a random force field with random
  annihilation/creation rates. The form taken by the  Lifshits
  singularity $\rho(E)\sim\exp-\frac{\pi|E|}{\sqrt{\sigma g}}$
  leads to the surprising conclusion that the average number
  of returning particles diverges at a finite time $t_c=\pi/\sqrt{g\sigma}$~:
  \begin{equation}
    \smean{n(x,t|x,0)} = \infty
    \hspace{0.5cm}\mbox{for}\hspace{0.5cm} t \geq t_c .
  \end{equation}
\end{itemize}
The two previous points show that this divergence of the {\it average}
particle density is due to the interplay between the random force field and
the random annihilation/creation rates. It would be an interesting issue to
understand precisely the physical origin of this remark.
On the other hand these last remarks might indicate that the white noise
$V(x)$ is probably too widely fluctuating for a reasonnable description of a
reallistic random annihilating/creating rates.
Maybe a more interesting model
would be to add a low concentration of such sites. In the continuum limit this
would correspond to add to the supersymmetric Hamiltonian a scalar potential
of the form $V(x)=\sum_n\alpha_n\delta(x-x_n)$, where $x_n$ are random
positions with a density $\rho$ and $\alpha_n$ local annihilation/creation
rates. The limit of high density $\rho\gg|\alpha_n|$ corresponds to the white
noise limit studied in the present article. The limit of low density
$\rho\ll|\alpha_n|$ might be more interesting. This model has been recently
studied in the absence of the random force field and for absorbing sites
($\alpha_n>0$) in Ref.~\cite{DeaSirSop06}, where a penetration length was
derived in any dimension thanks to renormalization group methods. An
interesting question would be to understand the effect of the random force
field on these known properties.


\section*{Acknowledgements}
CH is grateful to Pierre Le Doussal for his suggestion to study this model, in
particular for very useful discussions on the replica method. We would like to
Alain Comtet for valuable remarks and discussions.


\begin{appendix}

\section{The case of correlated noises\label{app:corrnoises}}

It is possible to extent the analysis to correlated noises in the following
sense. Suppose that $V(x) = \sqrt{\sigma}\eta(x)$ and
$\phi(x)=\mu\,g+\sqrt{g}\tilde{\eta}(x)$ are correlated such that
\begin{equation}
  \langle V(x)\phi(y)\rangle = \Gamma\delta(x-y)
\end{equation}
We introduce variables $\zeta(x) = \phi(x)-A$ and $v(x) =
2A(\phi(x)-\mu\,g)+V(x)$ so that the Hamiltonian may be rewritten as
\begin{equation}
  H = -\frac{\D^2}{\D x^2} + \zeta(x)^2+\zeta'(x)+v(x) + 2\mu g A-A^2
\end{equation}
The new variables have the correlation function
\begin{equation}
  \langle \zeta(x) v(y)\rangle = (2gA+\Gamma)\,\delta(x-y)
\end{equation}
so that the choice $A=-\Gamma/2g$ makes them independent. Further
characteristics are given by
\begin{align}
  \begin{array}{ll}
    \langle v(x)\rangle = 0, & \langle v(x)v(y)\rangle = \sigma\, \delta(x-y)\\
    \displaystyle\langle \zeta(x)\rangle = \mu\, g+\frac{\Gamma}{2g}, 
     & \langle \hat{\zeta}(x)\hat{\zeta}(y)\rangle = g \,\delta(x-y)
  \end{array}
\end{align}
where $\hat\zeta(x)=\zeta(x)-\mu\,g-\Gamma/2g$. Thus, up to a re-definition of
energy $\varepsilon = E +\mu\Gamma+\Gamma^2/4g^2$, we recover the problem of
uncorrelated noises.

\section{A useful relation}
\label{sec:usefulth}
Let us consider a random process generated by uncorrelated Wiener processes
$\D W_i(t)$~:
\begin{equation}
  \label{theorem}
  \boxed{
  \D x(t) = a(x)\, \D t + b_i(x)\,\D W_i(t)
  \eqlaw    a(x)\, \D t + \sqrt{b_i(x)b_i(x)}\,\D W(t)
  }
\end{equation}
The equality is valid for Ito and Stratonovich prescriptions.
Let us demonstrate this relation.

\vspace{0.25cm}

\noindent{\bf Ito's prescription.--}
Recall that the SDE 
\begin{equation}
  \D x_i = a_i(x)\, \D t + b_{ij}(x)\,\D W_j(t)
  \hspace{1cm}\mbox{  (Ito)}
\end{equation}
is associated to a FPE $\partial_tP=F_xP$ where the Forward Fokker-Planck
generator is \cite{Gar89}
\begin{equation}
  F_x = -\partial_ia_i + \frac12\partial_i\partial_jb_{ik}b_{jk}
\end{equation}
Therefore $\D x_i=a(x)\,\D t + b_{j}(x)\,\D W_j$ is associated to a FPE with
generator $F_x=-\partial_xa(x)+\frac12\partial_x^2b_{j}(x)b_{j}(x)$ that is
also associated to the SDE 
$\D x=a(x)\,\D t +\sqrt{b_i(x)b_i(x)}\,\D W(t)$. {\sc Qed}.

\vspace{0.25cm}

\noindent{\bf Stratonovich's prescription.--}
The relation between Ito and Stratonovich prescriptions is given in Ref.~\cite{Gar89}
\begin{eqnarray}
  \D x 
  &=& \alpha(x)\, \D t + \beta_{j}(x)\,\D W_j(t) \hspace{5cm} \mbox{(Stratonovich)} \\
  &=& \left[ \alpha + \frac12\beta_{j}\beta'_{j} \right]\D t
                       + \beta_{j}\,\D W_j(t) \hspace{4.15cm}\mbox{(Ito)} \\
  &\eqlaw& 
      \left[ \alpha + \frac12\beta_{j}\beta'_{j} \right]\D t
                       + \sqrt{\beta_{j}\beta_{j}}\,\D W(t) 
                      \hspace{3.5cm}\mbox{(Ito)} \\
  &=& \left[ 
          \alpha 
         + \frac12\beta_{j}\beta'_{j} 
         - \frac12\sqrt{\beta_{j}\beta_{j}}
           \left(\sqrt{\beta_{j}\beta_{j}}\right)'
      \right]\D t
                       + \sqrt{\beta_{j}\beta_{j}}\,\D W(t) 
  \hspace{0.25cm}\mbox{(Stratonovich)} \\
  &=& \alpha(x) \D t + \sqrt{\beta_{j}(x)\beta_{j}(x)}\,\D W(t) 
  \hspace{4cm}\mbox{(Stratonovich)}
\end{eqnarray}
{\sc Qed}.

This relation shows that addition law of variances holds not only for
additive processes but also for multiplicative processes.

\section{A remark on the Lyapunov exponent}
\label{sec:lyapunov}

In this appendix we clarify some relations between different formulae
for the Lyapunov exponent given above.

Let us present the problem with the well-known Halperin model
$H_\mathrm{scalar}=-\frac{\D^2}{\D x^2}+V(x)$. Here $V(x)$ denotes a
white-noise potential with average $\langle V(x)\rangle=0$, and
$\langle V(x)V(y)\rangle=\sigma \delta(x-y)$. The widely-used Ricatti
mapping allows to relate the spectral statistics for
$H_\mathrm{scalar}$ to passage probabilities for a diffusion $z(x)$
whose evolution is governed by the SDE $z'(x)=-[E+z(x)^2]+V(x)$. In
particular, the stationary distribution $T(z)$ for $z$ is solution to
the differential equation
\begin{equation}
\frac{\sigma}{2} T'(z) + (z^2+E)T(z) = {N}(E),
\label{eqn:statricatti}
\end{equation}
where $N(E)$ denotes the integrated density of states for
$H_\mathrm{scalar}$ as it can be shown from the node-counting theorem.
Moreover, the Lyapunov exponent relates to the diffusion via Rice
formula $\gamma = \langle z\rangle$ which, however, must be understood
as the principal value
\begin{equation}
  \gamma = \lim_{R\to +\infty}\int_{-R}^{R}\D z\, z\,T(z)
  \label{eqn:principal}
\end{equation}
in order to avoid difficulties from the asymptotic behaviour $T(z)\sim
N(E)/z^2$ as $|z|\to +\infty$. For $E>0$ \eqref{eqn:statricatti}
allows to rewrite
\begin{equation}
  \gamma=\lim_{R\to \infty}\int_{-R}^R \D z\,z\,
  \left(\frac{N(E)}{z^2+E}-\frac{\sigma}{2(z^2+E) }\,T'(z)\right)
\end{equation}
Clearly, the first term yields $0$. Notice that it is crucial to let
the integration bounds tend to $0$ symmetrically, otherwise we would
not find a well-defined result. After partial integration of the
second term, we eventually find an alternative expression for the
Lyapunov exponent
\begin{equation}
  \label{RelaLyap1}
  \gamma(E>0) 
  = -\frac{\sigma}{2}\int_{-\infty}^{+\infty} 
  \D z\,\frac{z^2-E}{(z^2+E)^2}\,T(z).
  = -\frac{\sigma}{2}\left\langle\frac{z^2-E}{(z^2+E)^2}\right\rangle
\end{equation}
The integration does not require anymore the principal value~: it was
possible to let the cutoff $R$ go to infinity since integrand now vanishes
sufficiently fast thanks to the partial integration.

This relation is particularly useful in order to study the limit
$E\to\infty$ since we may immediately read of the asymptotic behaviour
$\gamma\propto\sigma/E$. 
However the drawback is that (\ref{RelaLyap1})
is rather
ill-defined for $E\leq0$. Nevertheless, writing
\begin{equation}
  T(z) = \frac{{N}(E)}{z^2-E}-\frac{2E\,T(z)}{z^2-E}-\frac{\sigma\,T'(z)}{2(z^2-E)},
\end{equation}
it is not difficult to show that
\begin{equation}
  \label{RelaLyap2}
  \gamma(E<0)=-\left\langle\frac{2E z}{z^2-E}\right\rangle-\frac{\sigma}{2}\left\langle\frac{z^2+E}{(z^2-E)^2}\right\rangle
\end{equation}
by partial integration.  In order to extract the asymptotics, recall
that as $E\to -\infty$ the distribution $T(z)$ is centered at $z\sim
\sqrt{-E}$. Using this scaling behaviour we recover the asymptotic
behaviour $\gamma\propto\sqrt{-E}$.
It remains that the $E\to0$ limit in the two relations
(\ref{RelaLyap1},\ref{RelaLyap2}) seems tricky.

Let us now turn to our model Hamiltonian 
$H=-\frac{\D^2}{\D x^2}+\phi(x)^2+\phi'(x)+V(x)$. 
Section \ref{sec:ricatti} provides a
detailed account on the Ricatti mapping in this case, in particular
the stationary distribution $T(z)$ of the variable $z(x)$ was shown to
be solution of the differential equation
$N(E)=(z^2+E+2gz)T(z)+(\sigma+4gz^2)T'(z)/2$, see \eqref{eqT}. For
$E>0$ we may rewrite
\begin{equation}
  T(z)= \frac{N(E)}{z^2+E}-\frac{2g z\,T(z)}{z^2+E} - \frac{(\sigma+4gz^2)T'(z)}{2(z^2+E)}
\end{equation}
and insert this expression into \eqref{eqn:principal} what indeed allows to
recover \eqref{GammaRicatti}:
\begin{equation}
  \gamma(E>0) = \frac\sigma2
  \mean{
    \frac{E+\left(\frac{8Eg}{\sigma}-1\right)z^2}{(E+z^2)^2}
  }
  \label{eqn:gammaasymp}
\end{equation}
Conversely, for $E<0$ we may follow the same strategy as for Halperin's model
what yields an additional term
\begin{equation}
  \gamma(E<0) = -\left\langle\frac{2E z}{z^2-E}\right\rangle-\frac\sigma2
  \mean{
    \frac{E+\left(\frac{8Eg}{\sigma}+1\right)z^2}{(z^2-E)^2}
  }
\end{equation}
Again, the advantage of these formulae is that they provide the
asymptotic behaviour of $\gamma$ as $E\to\pm\infty$ in a very explicit
way. For example, as $E\to +\infty$ \eqref{eqn:gammaasymp} shows that
$\gamma\propto \sigma/E+4g$ what is coherent with $\gamma\sim
\gamma_\mathrm{susy}+\gamma_{\mathrm{scalar}}$.

\mathversion{bold}
\section{Solution of the differential equation $\mathcal{T}\Psi_0=0$}
\mathversion{normal}

\label{sec:sol}
The differential equation for $\Psi_0$ is given by
\begin{equation}
  \left(
    - \frac{\partial^2}{\partial\xi^2} 
    - \frac{E}{g^2} - \frac{\sigma}{4g^3}(\xi^2-1) + \frac{\mu^2-1/4}{\xi^2}
  \right)\Psi_0(\xi)=0
\end{equation}
In the absence of diagonal Gaussian disorder $\sigma=0$ we recover the
solution given in \cite{BouComGeoLeD90}. For $\sigma > 0$ we convert the
preceding equation into a differential equation for confluent hypergeometric
functions. Indeed, the ansatz
\begin{equation}
  \Psi_0(\xi) = \exp(\lambda \xi^2/2)\xi^\alpha w(z)\qquad\text{with }z=\eta\xi^2/2
\end{equation}
where $\alpha=1/2\pm \mu$, $\eta = -2\lambda = \pm \I\sqrt{\sigma/g^3}$ leads to 
\begin{equation}
  w''(z)+(b-z)w'(z)-aw(z)=0\qquad\text{with } b=\frac{1}{2}+\alpha,\,a=\frac{b}{2}-\frac{1}{2\eta}\left(\frac{E}{g^2}-\frac{\sigma}{4g^3}\right)
\end{equation}
The choice $\alpha=1/2+\mu$, $\eta=-2\lambda=+\I\sqrt{\sigma/g^3}$ leads to a
square integrable solution
\begin{equation}
  \Psi_0(\xi) = \exp\left(-\frac{\I}{4}\sqrt{\frac{\sigma}{g^3}}\,\xi^2\right)\xi^{\mu+1/2}\,U\left(\frac{\mu+1}{2}+\frac{\I}{2}\left(\frac{E}{\sqrt{g\sigma}}-\frac{1}{4} \sqrt{\frac{\sigma}{g^3}}\right),\mu+1,\I\sqrt{\frac{\sigma}{4 g^3}}\,\xi^2\right)
  \label{eqn:solpsi}
\end{equation}
In order to see square-integrability, recall that $E$ has a small negative
imaginary part $-i\epsilon$. Using
$U(a,b;z)\underset{z\to\infty}{\sim} z^{-a}+\dots$
we find
$|\Psi(\xi)|^2\underset{\xi\to\infty}{\sim}\xi^{-1-\epsilon}$. A second, linearly independent solution is readily found from complex conjugation of \eqref{eqn:solpsi}. The Wronskian which turns out to be useful for determination of the integrated density of states may be obtained from known properties of $U(a,b,z)$:
\begin{equation}
  W(\Psi_0(\xi),\overline{\Psi_0}(\xi))=2\I \exp (\pi\im a)\left(\frac{4g^3}{\sigma}\right)^{\mu/2}\xi^{1/2-\mu}\exp\left(\frac{\I}{4}\sqrt{\frac{\sigma}{g^3}}\xi^2\right)
  \label{eqn:wronskian}
\end{equation}  
\end{appendix}


\end{document}